\documentclass[10pt,aps,prd,twocolumn,nofootinbib]{revtex4-1}
 
\usepackage{amssymb,amsmath,accents}
\usepackage{aas_macros}
\usepackage{bm} % For \bm
\usepackage[shortlabels]{enumitem} % For enumeration with letters
\usepackage{subcaption,verbatim}%,subfig
\usepackage{soul} % for hl
\usepackage{graphicx}
\usepackage{color,units}
\usepackage{hyperref}
\usepackage{listings}
\usepackage[normalem]{ulem}
\hypersetup{
    unicode=false,          % non-Latin characters in Acrobat’s bookmarks
    pdftoolbar=true,        % show Acrobat’s toolbar?
    pdfmenubar=true,        % show Acrobat’s menu?
    pdffitwindow=false,     % window fit to page when opened
    pdfstartview={FitH},    % fits the width of the page to the window
    pdfauthor={Boris Goncharov},     % author
    colorlinks=true,       % false: boxed links; true: colored links
    linkcolor=blue,          % color of internal links
    citecolor=red,        % color of links to bibliography
}
\graphicspath{{plots_draft/}}

\begin{document}

\definecolor{dkgreen}{rgb}{0,0.6,0}
\definecolor{gray}{rgb}{0.5,0.5,0.5}
\definecolor{mauve}{rgb}{0.58,0,0.82}

\lstset{frame=tb,
  	language=Matlab,
  	aboveskip=3mm,
  	belowskip=3mm,
  	showstringspaces=false,
  	columns=flexible,
  	basicstyle={\small\ttfamily},
  	numbers=none,
  	numberstyle=\tiny\color{gray},
 	keywordstyle=\color{blue},
	commentstyle=\color{dkgreen},
  	stringstyle=\color{mauve},
  	breaklines=true,
  	breakatwhitespace=true
  	tabsize=3
}

\title{Utilizing the null stream of the Einstein Telescope} 

\author{Boris Goncharov}%
 \email{boris.goncharov@me.com}
\affiliation{Gran Sasso Science Institute (GSSI), I-67100 L'Aquila, Italy}
\affiliation{INFN, Laboratori Nazionali del Gran Sasso, I-67100 Assergi, Italy}

\author{Alexander H. Nitz}%
\affiliation{Max-Planck-Institut f\"ur Gravitationsphysik (Albert-Einstein-Institut), D-30167 Hannover, Germany}
\affiliation{Leibniz Universit\"at Hannover, D-30167 Hannover, Germany}

\author{Jan Harms}%
\affiliation{Gran Sasso Science Institute (GSSI), I-67100 L'Aquila, Italy}
\affiliation{INFN, Laboratori Nazionali del Gran Sasso, I-67100 Assergi, Italy}

\date{\today}

\begin{abstract}
Among third-generation ground-based gravitational-wave detectors proposed for the next decade, the Einstein Telescope provides a unique kind of null stream -- the signal-free linear combination of data -- that enables otherwise inaccessible tests of the noise models. We project and showcase challenges in modeling the noise in the 2030s and how it will affect the performance of third-generation detectors. We find that the null stream of the Einstein Telescope is capable of eliminating transient detector glitches that are known to limit current gravitational-wave searches. The techniques we discuss are computationally efficient and do not require \textit{a priori} knowledge about glitch models.
Furthermore, we show how the null stream can be used to provide an unbiased estimation of the noise power spectrum necessary for online and offline data analyses even with multiple loud signals in the band. We overview other approaches to utilizing the null stream. Finally, we comment on the limitations and future challenges of null-stream analyses for the Einstein Telescope and arbitrary detector networks.
%We demonstrate capabilities of the null stream to distinguish signals from transient detector glitches that are known to affect current gravitational-wave searches. In particular, we find that thanks to the null stream, \hl{the detection capability of the} Einstein Telescope will not be limited by glitches.
\end{abstract}

\maketitle

\section{Introduction}
The field of gravitational-wave (GW) astronomy, born with the first detection of a binary black hole (BBH) merger back in 2015~\cite{AbbottAbbott2016}, is only six years old. And yet, it has been very fruitful. The LIGO Scientific Collaboration and Virgo Collaboration have reported 90 GW signals, of which 3 are from the first observing run and 8 from the second observing run~\citep[GWTC-1,][]{AbbottAbbott2019b}, 44 signals from the first half of the third observing run~\citep[GWTC-2.1,][]{gwtc-2.1}, and 35 signals from the second half of the third observing run~\citep[GWTC-3,][]{gwtc-3}. In addition, an independent analysis has identified more than 94 mergers in these data~\citep[4-OGC,][]{NitzKumar2021}.
These signals have been observed with Advanced LIGO~\cite{LIGOScientificCollaborationAasi2015} and Advanced Virgo~\cite{AcerneseAgathos2015}, ground-based laser interferometers measuring a relative displacement of suspended test masses, which are affected by passing GWs with frequencies between $\sim 10 - 1000$\,Hz.
Moreover, pulsar timing arrays might be onto a detection of GWs between $\sim 10^{-9} - 10^{-8}$\,Hz with the reported noise process having a spectral slope similar to what is expected from the stochastic GW background from supermassive black hole binaries~\citep[e.g.][]{AntoniadisArzoumanian2022}. A number of other experiments are under development to probe the GW spectrum across around 20 orders of magnitude including the CMB B-mode polarization experiments \cite{BICEP2CollaborationAde2014} and the Laser-Interferometer Space Antenna~\citep[LISA,][]{Amaro-SeoaneAudley2017}.

Third-generation GW detectors (3G) -- the Einstein Telescope~\citep[ET,][]{PunturoAbernathy2010,ET2020} and the Cosmic Explorer~\citep[CE,][]{ReitzeAdhikari2019, EvansAdhikari2021} -- are ground-based laser interferometers expected to start operation in the mid 2030-s.  It is anticipated that third-generation observatories will detect $\approx 99.9\%$ stellar-origin binary black hole mergers in the Universe~\cite{MaggioreVanDenBroeck2020,KalogeraSathyaprakash2021}. However, amazing scientific prospects bring along new challenges. Signals will overlap in frequency and time and searches and data analysis pipelines will need to take that into account to ensure correctness and completeness of the results~\cite{SamajdarJanquart2021,PizzatiSachdev2021}. This will be especially relevant for ``low-latency'' searches that aim to detect and localize a binary neutron star (BNS) in the sky before the occurrence of its merger followed by a burst of electromagnetic radiation~\cite{AbbottAbbott2017a, CannonCariou2012, SachdevMagee2020, NitzSchafer2020, MageeChatterjee2021}. The detector noise power-spectral density is an important quantity in GW searches and is usually evaluated at time segments adjacent to a signal assumed to be signal free~\cite{VeitchRaymond2015} or using methods that either model~\cite{LittenbergCornish2015} or mitigate~\cite{UsmanNitz2016} only minor signal contamination. However, with improving sensitivity and especially at low frequencies there will be increasing contamination of the data from astrophysical sources. Signal-to-noise ratios will be reaching $\sim 1000$~\cite{HallEvans2019} for the brightest signals, which means that systematic errors in calibration will also need to remain below the order of $\sim 0.1\%$ in strain amplitude to not be significantly affecting the results. For a comparison, current systematic calibration errors are constrained to the $\approx 2\%$ level in amplitude~\citep{SunGoetz2020}.

\begin{figure}[!htb]
    \centering
    \includegraphics[width=1.0\columnwidth]{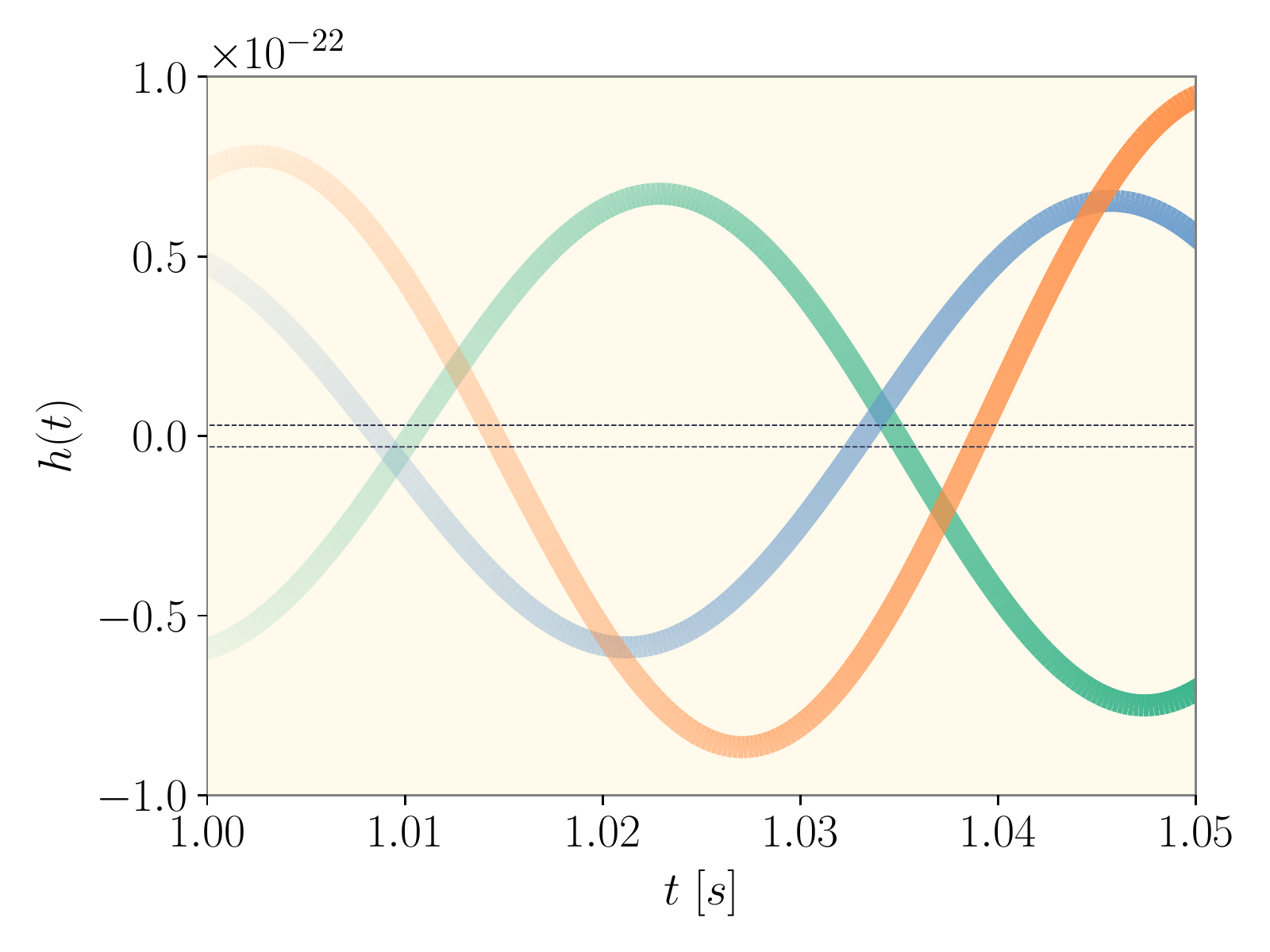}
    \caption{Demonstration of the null stream. Blue, green, orange lines represent the response of the Advanced LIGO at Hanford and Livingston, and the Advanced Virgo, respectively, to the GW signal. The linear null-stream combination of the detected signals, represented by the hollow dashed line, is zero.
    This network null stream is valid only for one signal because the linear coefficients depend on the source sky position.
    This is no longer the case for the ET null stream which is valid for any number of signals.
    }
    \label{fig:nullstreamdemo}
\end{figure}

The triangular design of the ET allows the construction of a unique signal-free data stream from the plain sum of strain output of constituent interferometers~\cite{FreiseChelkowski2009}.
This data stream is known as the \textit{null stream}.
Whereas it is still possible to construct a null stream for three arbitrarily located and oriented GW interferometers, it will only be applicable to one signal and will depend on its sky position~\cite{GurselTinto1989}. More generally, for $N_\text{det}$ GW detectors and $N_\text{pol}$ GW polarizations, it is possible to construct $(N_\text{det} - N_\text{pol})$ null streams.
%The requirement to have three interferometers follows from classical general relativity with two GW polarizations, referred to as \textit{plus} and \textit{cross}. \RED{JH: Where does this come from? I think that the network null stream does not depend on polarization, but only on propagation direction. N detectors can probably provide N-2 null streams. So, I believe that all the discussion about polarizations can be removed here.}
We provide demonstration of the null stream applied to a noiseless network of detectors in Figure~\ref{fig:nullstreamdemo}. Signals cancel in the null stream, but noise does not, which means that the null stream is a protection against any unmodeled instrumental noise.
This is precisely the class of problems we outlined in the previous paragraph. With that, thanks to the closed triangular geometry and negligible light travel times between constituent interferometers, the null stream in ET is not attached to a specific signal, and thus works independently of the GW propagation direction and therefore also in the case of many GW signals present at the same time. This makes the ET null stream invaluable to a range of data-analysis problems. The ET null stream is a rare experimental opportunity to test if one's noise models are correct, which is often not true and historically caused several spurious detections in physics~\citep[e.g.][]{AdamAgafonova2012,PlanckCollaborationAdam2016}.

The first example of employing a null stream for arbitrary detector networks was brought forward by G\"ursel and Tinto~\cite{GurselTinto1989}. They suggested a reconstruction of source direction and GW strain time series from an arbitrary signal through minimizing residuals in the null stream. Wen and Schutz~\cite{WenSchutz2005} pointed out that vetoing signals based on the null stream can supersede coincidence-based vetoes while protecting coherent methods from confusion by spurious noise. Ajith, Hewitson and Heng~\cite{AjithHewitson2006} assessed the effect of uncertainties in detector calibration in the null-stream veto analysis. A more detailed mathematical description of addressing these problems is provided in Chatterji et al~\cite{ChatterjiLazzarini2006}. 
Rakhmonov~\cite{Rakhmanov2006} pointed out that the G\"ursel-Tinto problem belongs to the class of ill-posed problems because of rank deficiency of the network detector response matrix and demonstrated the Tikhonov regularization procedure. Harry and Fairhurst~\cite{HarryFairhurst2011} introduced null SNR and distinguished two approaches to constructing the null stream for signal consistency tests.
In the first ET mock data challenge, Regimbau et al.~\cite{RegimbauDent2012} discuss applications of the null stream and show how it can be used to estimate the average power spectrum of astrophysical signals.
Schutz and Sathyaprakash~\citep{SchutzSathyaprakash2020} suggested that residuals in the null stream can be used for self-calibration of detector networks. Wong et al.~\cite{WongPang2021} developed a null stream framework to search for generic GW polarizations. Wong and Li~\cite{WongLi2022} focused specifically on ET and showed how to reduce dimensionality of a three-detector likelihood to an effectively two-detector likelihood based on the null-stream redundancy. The null stream is also utilized in other GW experiments. For pulsar timing arrays, Zhu et al.\;suggested to utilize null streams in continuous GW searches~\cite{ZhuWen2015,ZhuWen2016}. Goldstein et al.~\citep{GoldsteinVeitch2018} extended the methodology to other signals while combining it with Bayesian statistics and pointing out a different way to construct null streams. We also point out that the LISA T channel \cite{PrinceTinto2002} corresponds to the ET null stream. It is a more accurate description of the null stream taking light travel times across the detector into account, which are neglected in the ET null-stream formalism. 
The most recent work by Janssens et al.~\cite{JanssensBoileau2022} addresses the case of non-identical noise, both correlated and uncorrelated, in the context of LISA and ET.

In this paper, we assess the role of the null stream for various data analysis problems with 3G detectors.
A brief overview of the relevant aspects of the ET design is given in Section \ref{sec:ET}. In Section~\ref{sec:method}, we explain the construction and application of null stream for ground-based laser interferometers.
In Sections~\ref{sec:glitches} and~\ref{sec:psd}, we discuss the problems of separating signals from glitches and estimating noise power spectra in ET, respectively, along with the solutions provided by the ET null stream.
We discuss other applications and limitations of null streams in Sections~\ref{sec:other} and~\ref{sec:limitations}, respectively.
The summary and the conclusions are provided in Section~\ref{sec:conclusion}.

% (perhaps emphasize here also how this is different than what can be done now?)
% summarize the impacts this has: 
%
%    Precision [more importantly tight control on systematics] GW measurements
%           estimation / modeling of calibration errors
%           estimation / modeling of PSD (sans confusion noise)
%           inclusion of glitch model in PE
%           (could summarize these as access to direct channel that characterizes any mismodeling of the signal or noise properties)
%
%    Clean Detections due to glitch removal / veto
%    Tests of non-GR polarizations
%    stochastic background 

\section{The Einstein Telescope}
\label{sec:ET}
The Einstein Telescope (ET) is a proposed underground infrastructure to host next-generation GW detectors \cite{PunturoAbernathy2010,ET2011,ET2020}. Its geometry is an equilateral triangle with 10\,km side length. The structure will host 6 laser interferometers in a so-called xylophone configuration \cite{HildChelkowski2010}, which combines a low-frequency (LF) optimized interferometer with a high-frequency (HF) optimized interferometer for each of the three vertices. 

We will refer to an LF/HF interferometer pair as ET \emph{component}. Adding data from the three (accurately calibrated) components provides a null stream \cite{GurselTinto1989}. An underlying assumption is that all components operate in parallel.  However, since there are a total of six interferometers, each with their own duty cycle, it is clear that the time when all interferometers operate simultaneously is reduced. This means that the ET null stream cannot be formed at all times.

Another important aspect of some applications of the null stream is noise correlations between ET components. Noise correlations were a major concern at LIGO Hanford in the past when it still operated two collocated interferometers \cite{AbbottAbbott2009}. These two interferometers shared the same vacuum system, which might have been one of the key issues. Also, some of the test masses of the two LIGO Hanford detectors were located close to each other. The ET interferometers will all be located inside separate vacuum systems~\citep{ET2020}, which should strongly reduce noise correlations between interferometers. Furthermore, test masses of different ET interferometers are separated by at least a few hundred meters. Still, it is conceivable that magnetic fluctuations, or gravitational fluctuations from ambient seismic and acoustic fields produce significant noise correlations. It is in principle possible to monitor the respective environmental fields and subtract this noise from data. However, such cancellation schemes have only been demonstrated on rather simple problems so far \citep{DriggersVitale2019}, and their effectiveness in ET cannot be accurately predicted today. 

\section{\label{sec:method} Methodology}
The null stream is a synthetic data channel, which cannot provide more information on the GW signal than the original data it is based on. However, it is possible to use the null stream to test noise models, which are not known \textit{a priori}. Detector noise varies slowly in time, and it contains short transients known as glitches \cite{DavisAreeda2021}. Analyzing GW data requires correct noise models, and this is where the null stream gets extremely valuable. In this Section, we outline the process of application of the null stream to unspecified or misspecified noise processes. Keeping in mind several existing methods~\cite{GurselTinto1989,WenSchutz2005,AjithHewitson2006,ChatterjiLazzarini2006,Rakhmanov2006,HarryFairhurst2011} of employing null streams in analyses of data from ground-based interferometers, we outline the construction of the null stream in Section~\ref{sec:null_stream}, the likelihood-based approach to the null stream in Section~\ref{sec:likelihood}, and the approach to estimate the noise power spectrum with the null stream in Section~\ref{sec:psd_method}. We emphasize the role of the null stream in the Einstein Telescope, which is significantly more versatile than the general null stream and simpler to apply in practice.

\subsection{\label{sec:null_stream} Null stream construction}
Gravitational-wave detectors have intrinsic noise, $\bm{n}_i$, while the GW signals projected onto detectors add their contributions $\bm{h}_i$ on top of the noise.
Following G\"ursel \& Tinto~\cite{GurselTinto1989}, for any three GW strain responses $\bm{h}_{1,2,3}$, we find that the signal in one detector can be expressed as a linear combination of signals in the other two detectors according to:
\begin{equation}\label{eq:ns_strain}
    h_1(t|\bm{\Omega}) = \eta(\bm{\Omega}) h_2(t + \tau_{12}) + \zeta(\bm{\Omega}) h_3(t+\tau_{13}),
\end{equation}
where $\tau_{ij}$ are time delays between detectors, and $\eta(\bm{\Omega})$ and $\zeta(\bm{\Omega})$ are combinations of the GW antenna response functions $F^{1,2,3}_{+,\times}(\bm{\Omega})$
\begin{equation}\label{eq:ns_coeffs}
    \begin{cases}
        \eta(\bm{\Omega}) = -\cfrac{F^3_\times F^1_+ - F^3_+ F^1_\times}{F^2_\times F^3_+ - F^2_+ F^3_\times}, \\[10pt]
        \zeta(\bm{\Omega}) = -\cfrac{F^1_\times F^2_+ - F^1_+ F^2_\times}{F^2_\times F^3_+ - F^2_+ F^3_\times}.
    \end{cases}
\end{equation}
The coefficients $\eta,\,\zeta$ as well as the time delays $\tau_{ij}$ only depend on the propagation direction $\bm{\Omega}$ of the GW.

Since signals in three detectors are linearly dependent, we can define the null stream data as a linear combination of time-shifted detector strain measurements (data) $\bm{d}_i = \bm{n}_i + \bm{h}_i$ where $\bm{h}_i$ completely cancel each other out following equation (\ref{eq:ns_strain}):
\begin{equation}
\begin{split}
    d_\text{null}(t|\bm{\Omega}) \equiv n_\text{null}(t|\bm{\Omega}) = \\ d_1(t) - \eta(\bm{\Omega}) d_2(t + \tau_{12}) - \zeta(\bm{\Omega}) d_3(t+\tau_{13}).
\end{split}
\end{equation}
Thus, the null stream data contain only the noise and no GW signals.
One may notice that the three equations above represent the cross-product of $\bm{F}_+ = (F^1_+, F^2_+, F^3_+)$ and $\bm{F}_\times = (F^1_\times, F^2_\times, F^3_\times)$, which means that the null stream is orthogonal to the space of $F_+$ and $F_\times$~\citep[see Equation 17 in][]{ChatterjiLazzarini2006}. % One can also add that cross product is the determinant of the matrix, and null stream of zero means that the components are linearly dependent.
Clearly, the null stream can be formed in time domain as well as in frequency domain, where time delays turn into complex phases. For a generic detector network, the null stream for one signal will contain residuals of the other signals if they are simultaneously present in the data and not coming from the same sky location. 

For the special case of the Einstein Telescope (ET) made of three almost co-located interferometers with arms forming an equilateral triangle, $\eta=\zeta=-1$ and $d_\text{null}(t)$ does not depend on the sky position.
In ET, the effective number of vectors along the arms of ET components, that make up $\bm{F}_{+,\times}$, is reduced to three, which cancels nominators and denominators in Equation~\ref{eq:ns_coeffs}~\citep[see Section 3.2 in][]{SathyaprakashAbernathy2012}.
Therefore, any combination of GW signals cancels in the ET null stream. The only condition for this null stream to hold is that two parallel and equally long ET interferometer arms experience the same length change due to a GW.
It entails that time delays between the three detector components are negligibleIn a more general case, suppression of GW signals in the ET null stream becomes less efficient towards high frequencies similar to the case of the T-channel combination of the LISA detector \citep[see Figure 5 in][]{PrinceTinto2002}.
The residuals in the null stream are to be proportional to the wave amplitude change during the time lag, $\approx 2 \pi f L/c$, as well as the numerical factor that depends on the sky position and polarization.
%\hl{With the gravitational wave amplitude change $\Delta h \approx \sin(2 \pi f L /c) \approx 2 \pi f L/c$ during the time delay $L/c$, as well as the sky position and polarization dependent factor $<0.25$, we roughly estimate the residuals in the null stream due to be $ \lessapprox 5 \times 10^{-5} f h$.}

\subsection{\label{sec:likelihood} Null-stream likelihood}

Assuming that the noise $\bm{n}_i$ follows a Gaussian distribution, we employ the Gaussian likelihood~\cite{AshtonHubner2019} to describe the data $\tilde{\bm{d}}_i$ of the detector $i$,
\begin{equation}\label{eq:likelihood}
\begin{split}
\log \mathcal{L}(\tilde{\bm{d}}_i|\bm{\theta}) \sim -\frac{1}{2}\langle \tilde{\bm{d}}_i - \tilde{\bm{h}}_i | \tilde{\bm{d}}_i - \tilde{\bm{h}}_i\rangle,
\end{split}
\end{equation}
where $\bm{\theta}$ are parameters for our signal and noise models, the tilde denotes a Fourier transform, and the brackets represent the noise-weighted inner product:
\begin{equation}\label{eq:inner_product}
\begin{split}
\langle \tilde{a}|\tilde{b}\rangle=4\Delta f \sum\limits_{k=1}^{N_\text{f}}\frac{\Re(\tilde a(f_k)\tilde b^*(f_k))}{P(f_k)}. 
\end{split}
\end{equation}
The asterisks denote complex conjugation.
The integral of the likelihood for a given model over the prior probability $\pi(\bm{\theta})$ is referred to as the evidence, $\mathcal{Z}$.
We calculate it by means of nested sampling~\cite{Skilling2004,HandleyHobson2015a,HandleyHobson2015b}.
The Bayes factor $\mathcal{B}$ is the ratio of evidences for two models, and it is used to determine which model of the two best fits the data.
The sum is over the $N_\text{f}$ frequency bins of width $\Delta f=1/T$ used for the discrete Fourier transform.
The power spectral density (PSD) of the detector noise, $\bm{P}$, is~\cite{ThorneBlandford2017}:
\begin{equation}
\label{eq:psd_T}
    P(f) = \lim_{T \xrightarrow{} \infty} \frac{2 |\tilde{d}(f)|^2}{T}.
\end{equation}
Usually, $\bm{P}_i$ -- the noise PSD of detector $i$ -- is determined empirically from the data prior to the GW event~\cite{ChatziioannouHaster2019}.
The likelihood, through the inner product, implicitly includes a diagonal covariance matrix for frequency bins, with elements $\bm{\sigma}^{-2}_i = 4 \Delta f / \bm{P}_i$.
If the noise in the three components of ET is uncorrelated, we can write the total likelihood as a product of individual likelihoods,
\begin{equation}\label{eq:likelihood_product}
    \mathcal{L}(\tilde{\bm{d}}|\bm{\theta}) = \prod_{i=1}^{N_\text{det}} \mathcal{L}(\tilde{\bm{d}}_i|\bm{\theta}),
\end{equation}
where, for ET, $N_\text{det} = 3$ and $\tilde{\bm{d}} = (\tilde{\bm{d}}_1, \tilde{\bm{d}}_2, \tilde{\bm{d}}_3)$. Similarly, if the noise is stationary and the measurements are continuous, the noise in different frequency bins $f_j$ is independent and thus a likelihood can be written as a product of likelihoods for each of the $N_\text{f}$ frequency bins: $\mathcal{L}(\tilde{\bm{d}}_i|\bm{\theta}) = \prod_{j=1}^{N_\text{f}} \mathcal{L}(\tilde{d}_{i,j}|\bm{\theta})$.
Note, in ET, due to null stream redundancy for any number of signals, the product in Equation~\ref{eq:likelihood_product} can be reduced to have $(N_\text{det} - 1)$ effective terms~\cite{WongLi2022}.

It is also possible to introduce a likelihood term for the null stream,
\begin{equation}\label{eq:likelihood_null}
   \log \mathcal{L}(\tilde{\bm{d}}_\text{null}) \sim -\frac{1}{2}\langle \tilde{\bm{d}}_\text{null} | \tilde{\bm{d}}_\text{null} \rangle.
\end{equation}
Because it contains only the Gaussian noise model, in our case parameter-free, $\log \mathcal{L}(\bm{d}_\text{null})$ also represents Bayesian evidence of Gaussian noise in the null stream. If the noise is modeled correctly and is therefore cancelled in the null stream, $\log \mathcal{L}(\tilde{\bm{d}}_\text{null}) \approx -N_\text{f}$, which follows from Equations~\ref{eq:likelihood},~\ref{eq:inner_product},~\ref{eq:psd_T}.
%\hl{(isn't it true from the likelihood definition that the null-stream log-likelihood, if it is a correctly modeled Gaussian process, has an (average) value of $-2N_\text{f}$? of course, this is just an arbitrary normalization, but we should either stick to the definition in eq (4), or briefly say that we renormalize the likelihood to divide out the (expected) Gaussian noise term)}.
For $\log \mathcal{L}(\tilde{\bm{d}}_\text{null}) < -N_\text{f}$, the Kass-Raftery scale~\citep{kassraftery} can be used to quantify evidence for improperly modeled noise.

%Because the linear combination of a Gaussian random variable is another Gaussian random variable, $\bm{d}_i$ are independent with the null-stream $\bm{d}_\text{null}$ because $\bm{n}_i$ are independent \hl{(actually, I believe that this sentence is incorrect. the noise of the detector i is correlated with the null stream)}. Thus, the null stream is an independent Gaussian data stream with the noise variance $\bm{\sigma}^2_\text{n}$ represented by the sum of $\bm{\sigma}_i^2$ multiplied by $\eta$ and $\zeta$ where appropriate.
%Given the independence condition, we can combine the null stream likelihood term with the total likelihood \hl{(I believe that this is incorrect for the same reason as pointed out before. the data in dnull and the detector data are not independent! anyway, I think that we can leave out this entire last paragraph)}:
% \begin{equation}
% \label{eq:nslnltotal}
%     \mathcal{L}(\bm{d},\bm{d}_\text{null}|\bm{\theta}) = \mathcal{L}(\bm{d}_\text{null}) \prod_{i=1}^{N} \mathcal{L}(\bm{d}_i|\bm{\theta}).
% \end{equation}

\subsection{\label{sec:psd_method} Null stream and noise power spectrum}

The null stream of ET enables an elegant method to provide unbiased estimates of instrument noise power spectral densities (PSDs) in low latency. The PSD $\bm{P}_i$ of the detector component $i$ can be calculated as a cross-spectral density between the null stream $\bm{d}_\text{null}$ and the detector data $\bm{d}_i$,
\begin{equation}
\bm{P}_i=\bm{d}_\text{null}\star\bm{d}_i.
\label{eq:noisepsd}
\end{equation}
An asterisks denotes a convolution operator for calculating the cross-correlation.
This equation follows from the assumption that noise is uncorrelated, such that $\bm{d}_i\star\bm{d}_j = 0$ when $i \neq j$. For example, assuming that $\bm{n}_1$ and $\bm{n}_2$ are correlated, e.g., due to fluctuations of the magnetic field \cite{ThraneChristensen2013}, we can write $\bm{n}_2$ as a linear combination of $\bm{n}_1$ and a new independent noise $\bm{n}_2'$,
\begin{equation}
\bm{n}_2 = a(f)\bm{n}_1+\bm{n}_2',
\end{equation}
where the complex coefficient $a(f)$ and the PSD of $\bm{n}_2'$ are determined by the cross PSD between $\bm{n}_1$ and $\bm{n}_2$ and by the PSDs $\bm{P}_1,\,\bm{P}_2$. Now, the cross PSD between the null stream $\bm{d}_\text{n,corr}$ affected by noise correlations and $\bm{d}_1$ becomes
\begin{equation}
    \bm{d}_\text{n,corr}\star\bm{d}_1=(1+a)\bm{P}_1,
\end{equation}
which means that the estimate is biased by the correlation coefficient $a$ between the two ET components. It is not possible to estimate the coefficient $a$ from other measurements in the presence of a significant GW background. One can ameliorate this problem only by anticipating noise correlations and suppressing them, e.g., using noise-cancellation techniques \cite{CoughlinCirone2018,BadaraccoHarms2019}. Additionally, if noise between detectors is uncorrelated, the power spectral density of the null stream can be represented as:
\begin{equation}\label{eq:psd}
    \bm{P}_\text{n} = \bm{P}_1 + \eta^2 \bm{P}_2 + \zeta^2 \bm{P}_3,
\end{equation}
where $\eta = \zeta = -1$ for ET, as discussed above.
Throughout this work, we simulate detector noise based on the design sensitivity of the ET in the past referred to as ET-D~\cite{HildAbernathy2011}.

\section{\label{sec:glitches} Separating signals from glitches}
%\RED{There might be basic/generic parts in this section that might be better to have in section \ref{sec:method}.}

\begin{figure*}[!htb]
    \centering
    \begin{subfigure}[b]{0.49\textwidth}
        \includegraphics[width=\textwidth]{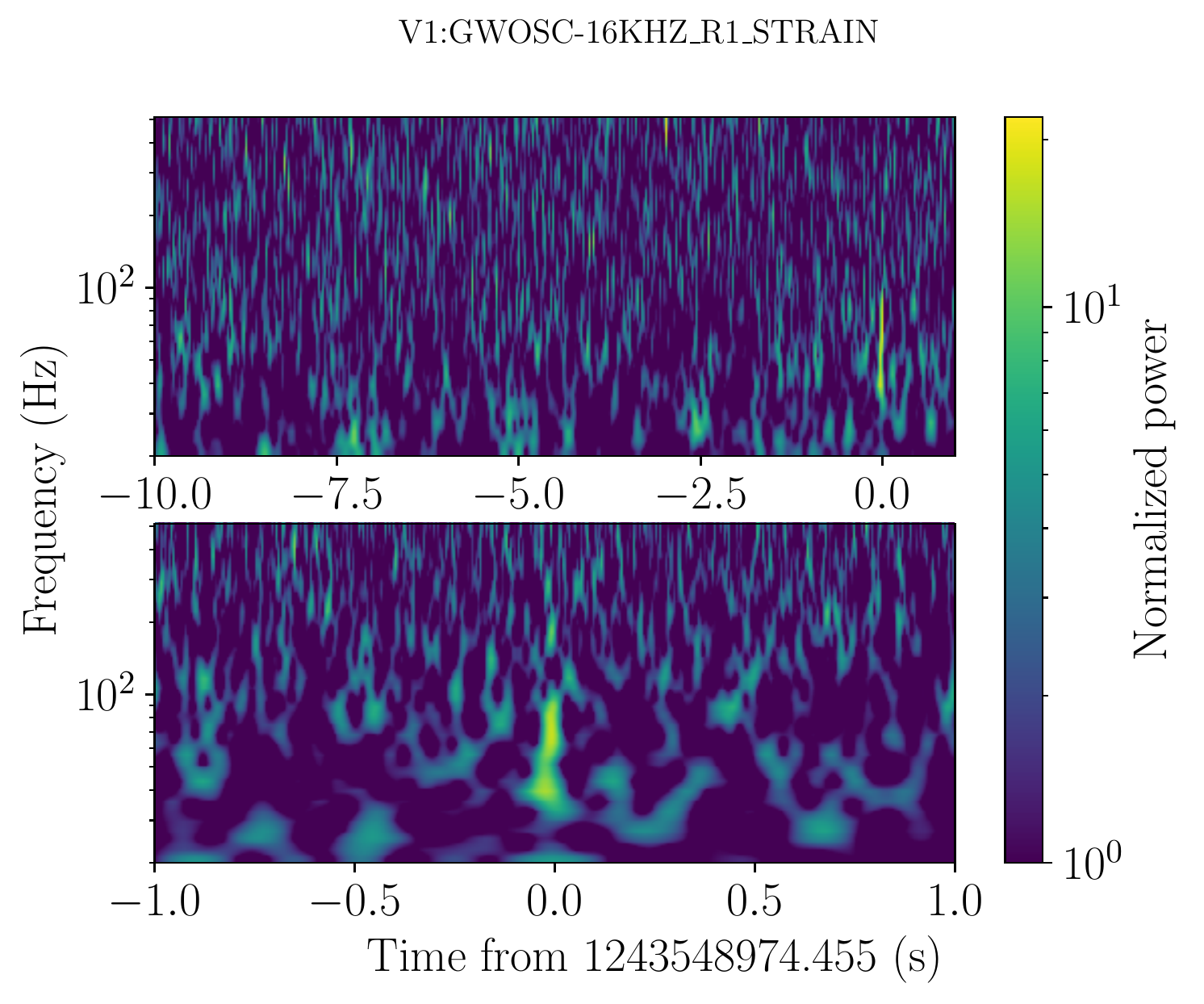}
      \caption{Glitch}
        \label{fig:snrs:signal}
    \end{subfigure}
    \begin{subfigure}[b]{0.49\textwidth}
        \includegraphics[width=\textwidth]{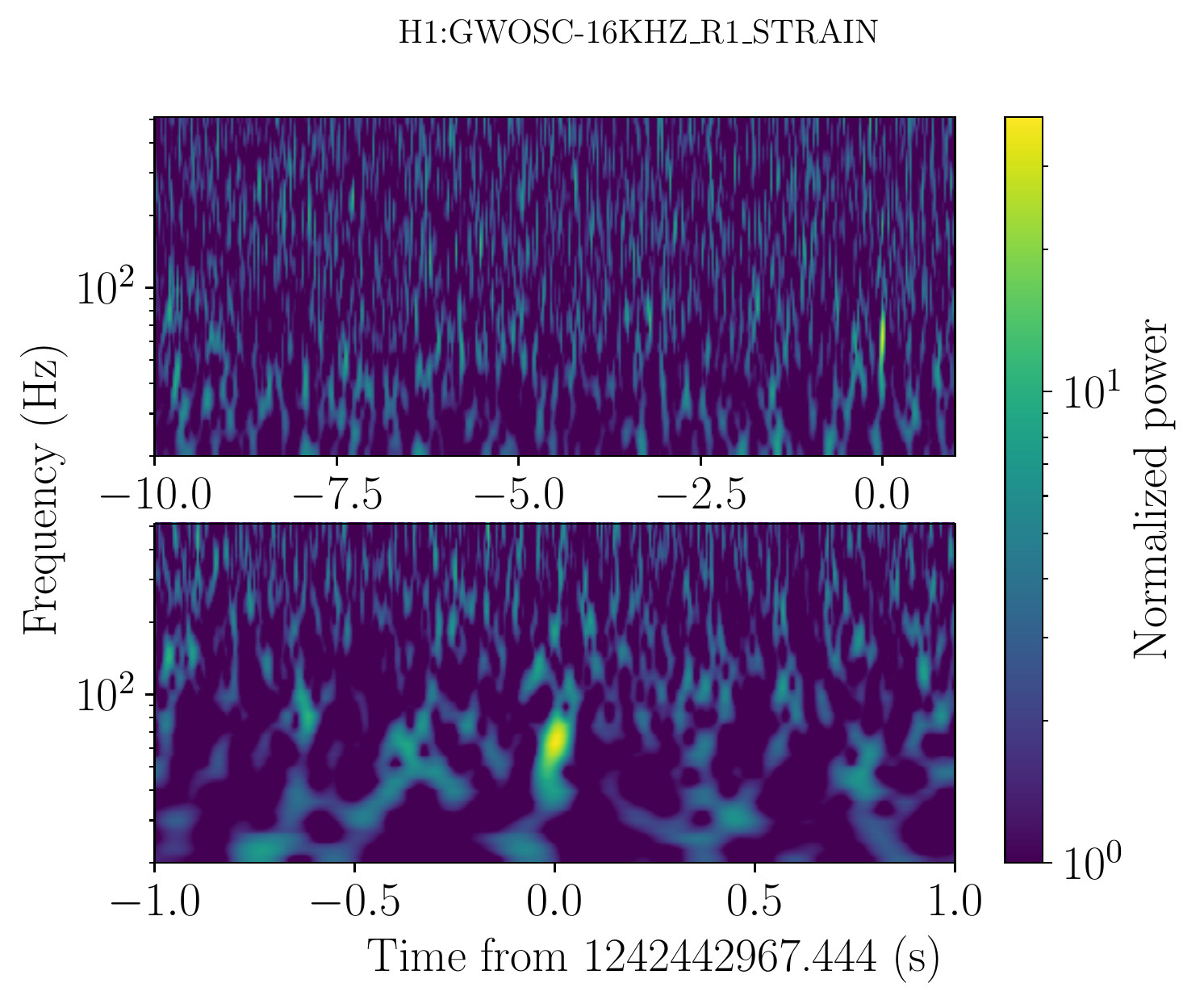}
      \caption{GW190521}
        \label{fig:snrs:glitch}
    \end{subfigure}
    
    \caption{Visual similarity between a compact binary merger signal and the instrumental artifact known as the blip glitch. The panels represent time-frequency diagrams of the normalized power in the detectors. Left: a glitch observed by the Virgo instrument. Right: GW190521~\cite{AbbottAbbott2020}, a binary black hole merger with a total mass of $150 M_\odot$. We consider such high-mass high-redshift events as one class of the signals that are difficult to distinguish from the noise, especially when these signals are weak and when detector sensitivities differ substantially.
    }
    \label{fig:glitchsignal}
\end{figure*}

Current GW observatories suffer from a number of instrumental noise artifacts; these can be stationary and non-stationary. Stationary sources of noise include alternating current power harmonics and repeating instrumental lines with various frequency spacing known as combs~\cite{CovasEffler2018}.
These stationary sources mostly affect searches for stochastic or continuous GWs. Non-stationary sources of noise include glitches~\cite{CaberoLundgren2019} and simply time segments with bad data~\cite{DavisAreeda2021}, and generally contaminate all GW searches. Glitches affect the performance of online searches, with the most well-known example when a blip glitch in the LIGO Livingston detector has happened during the BNS merger GW170817~\cite{AbbottAbbott2017b}, thus delaying the initial sky localizations for electromagnetic follow-up~\cite{AbbottAbbott2017a}. While approaching the design sensitivity, LIGO and Virgo not only discover more signals, but also retract more apparent signals~\cite{CaberoMahabal2020}. We expect this trend to continue towards the construction and operation of 3G detectors. Moreover, so-called blip glitches have a similar morphology to mergers of high-mass high-redshift binary black holes (BBHs) ~\cite{CaberoLundgren2019} and roughly similar peak frequencies~\cite{Nitz2018}. An example is shown in Figure~\ref{fig:glitchsignal}.

Furthermore, for certain signals we only have phenomenological models and distinguishing these from non-stationary noise is more challenging. These can be burst signals from supernovae or cosmic strings~\cite{AbbottAbbott2021b}, which are modeled as, e.g., a superposition of sine-Gaussians~\cite{DragoKlimenko2021} because of the complex physics and time-frequency dependence of the waveform. Another example are continuous GW signals from neutron stars with unknown frequency evolution from spin wandering~\cite{TheLIGOScientificCollaborationtheVirgoCollaboration2022,SuvorovaSun2016} or other sources of persistent GWs, which are considered stochastic~\cite{AbbottAbbott2019a,AbbottAbbott2021a,GoncharovThrane2018}. Whereas there are several consistency tests for such signals~\cite{Allen2005,Nitz2018,ChatziioannouCornish2021,GoncharovThrane2018, NitzDent2017,CaberoLundgren2019, McIsaacHarry2022}, there are examples where certain outliers were reported without clear picture of whether their origin is astrophysical or instrumental~\citep[e.g.,][]{AbbottAbbott2019a,DergachevPapa2020}. For short-duration sources, even though the gravitational-wave signal can be well modeled, it is often challenging to confirm the astrophysical origin~\cite{CaberoLundgren2019}. 

The null stream of the Einstein Telescope enables a complementary glitch veto suitable for the cases described in this paragraph.
%The advantage of this test is that it is independent of glitch and signal models and provides an unambiguous result.
%\hl{(I reformulated this paragraph slightly. still fine?)}
We suggest two straightforward implementations in the following subsections.
The first approach depends on the signal model and the second approach is independent of it.
Both approaches are based on assumptions about noise properties of the null stream that we outlined in previous paragraphs.

\begin{figure*}[!htb]
    \centering
    \begin{subfigure}[b]{0.32\textwidth}
        \includegraphics[width=\textwidth]{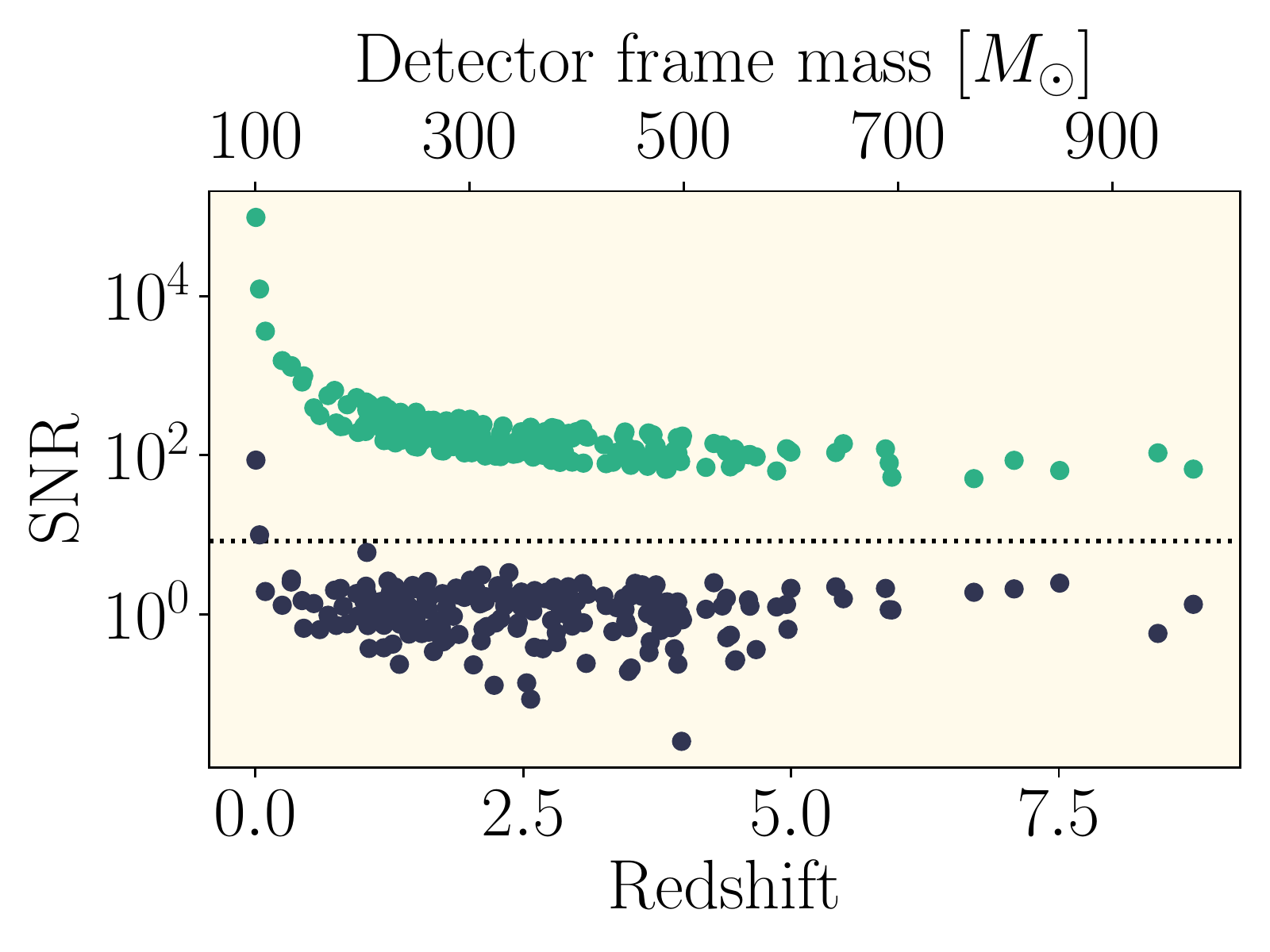}
      \caption{Field ($z \sim 2$)}
        \label{fig:snrs:1}
    \end{subfigure}
    \begin{subfigure}[b]{0.32\textwidth}
        \includegraphics[width=\textwidth]{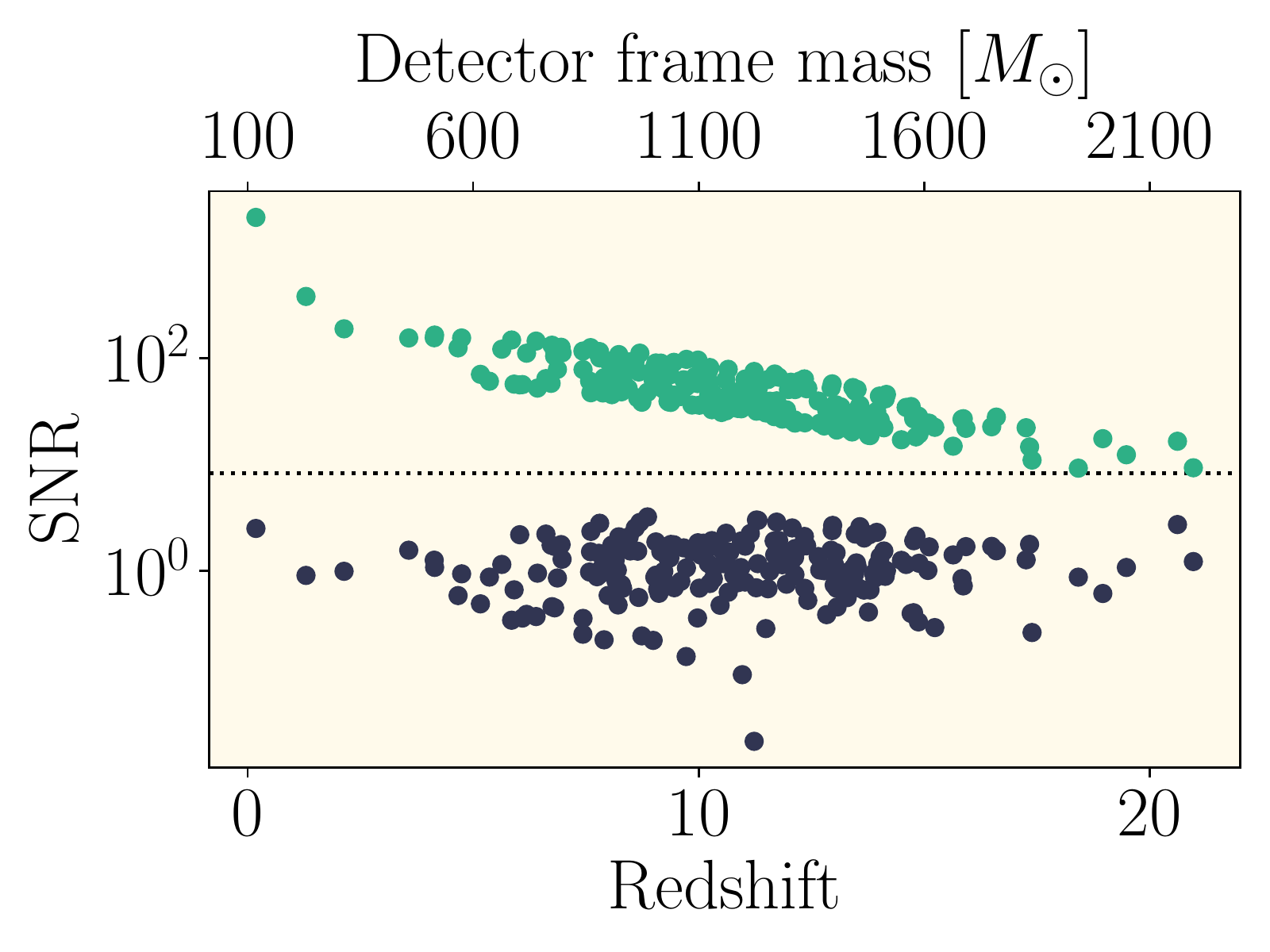}
      \caption{Population-III ($z \approx 10$)}
        \label{fig:snrs:2}
    \end{subfigure}
    \begin{subfigure}[b]{0.32\textwidth}
        \includegraphics[width=\textwidth]{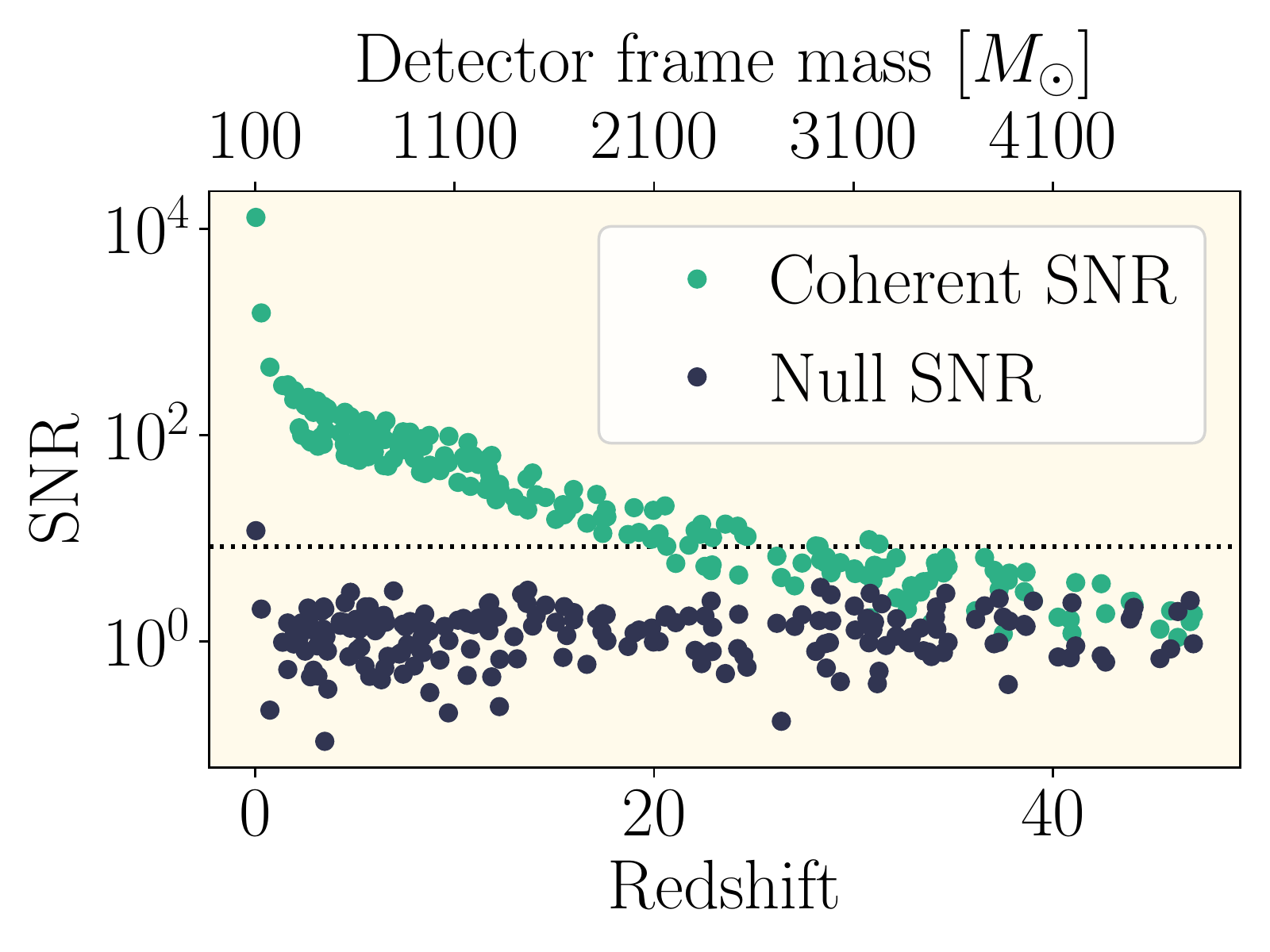}
        \caption{Primordial}
        \label{fig:snrs:3}
    \end{subfigure}
    \caption{Demonstrating the cancellation of GWs from BBH mergers in the null stream of Einstein Telescope. We consider equal-mass binary black holes with source-frame component masses of $100~M_\odot$. Three panels correspond to three fiducial source redshift distributions.
    Event redshifts are on the bottom horizontal axes and detector-frame masses are on the top horizontal axes. Vertical axes represent SNR and the horizontal dotted line corresponds to a detection threshold of SNR $=8$. Green points represent coherent SNR quantifying the significance of the detection. Dark-blue points represent null SNR for the same events. Null SNR can be employed as a veto for glitches and other non-stationary noise. A few loud close-by signals at $z \approx 0$ and coherent SNR $> 10^4$ have significant null SNR because of neglected time delays between ET components. The signals at high redshifts with low SNR and short duration are the ones that may be the primary candidates for null stream follow-up, because of their similarity to glitches, as shown in Figure~\ref{fig:glitchsignal}. These are primarily population-III and primordial BBHs.}
    \label{fig:snrs}
\end{figure*}

%Systematic errors. $d = s + n$, $s = h + g$. $N d = 0 + N g + n$. 

\subsection{\label{sec:glitches:nullsnr} Null SNR and BBH populations}

The first implementation is to introduce null SNR from Harry and Fairhurst~\cite{HarryFairhurst2011}:
\begin{equation}
\begin{split}
    \rho_\text{null}^2 = \rho_\text{coinc}^2 - \rho_\text{coh}^2 = \\ \sum_{X,Y} \sum_{\phi=0,2\pi} \frac{\langle d_X | h_\phi \rangle}{\sigma_X} N^{XY} \frac{\langle d_Y | h_\phi \rangle}{\sigma_Y},
\end{split}
\end{equation}
where
\begin{equation}
\begin{cases}
\begin{aligned}
    & N^{XY} \equiv \delta^{XY} - (f_+^X f_+^Y + f_\times^X f_\times^Y),\\[0.3ex]
    & f^{X,Y}_{+,\times} \equiv \dfrac{\sigma^{X,Y}F^{X,Y}_{+,\times}}{\Sigma_{Z} (\sigma^{Z} F_{+,\times}^{Z})^2}, Z \neq X, Y,\\[0.3ex]
    & \sigma^{X,Y} \equiv \sqrt{\langle h_0^{X,Y} | h_0^{X,Y} \rangle} \approx \sqrt{\langle h_{\pi/2}^{X,Y} | h_{\pi/2}^{X,Y} \rangle},
\end{aligned}
\end{cases}
\end{equation}
$h_{0,\pi/2}$ are the phases of the waveform, $X$ and $Y$ are detectors.
Null SNR is the difference in quadrature between the coincident SNR, $\rho_\text{coinc}$, and the coherent SNR, $\rho_\text{coh}$.
Ultimately, null SNR represents the coincident SNR in the null stream for the case of equal detector component sensitivities.
We discuss an implementation of null SNR for the Einstein Telescope using \texttt{IMRPhenomD} gravitational waveforms~\cite{KhanHusa2016,HusaKhan2016}.
This is the standard choice for frequency-domain analyses when precession and high-order modes are not of interest.

First, we demonstrate the impact of astrophysical signals on coherent SNR and null SNR.
We inspect black hole merger signals with $100~M_\odot$ source-frame masses from three fiducial populations of compact binaries distributed isotropically across the sky. We choose such a high source-frame mass to focus on the short-duration signals, which will have a similar morphology and duration to glitches.
To be precise, in this section, we discuss field binaries with the volumetric merger rate [$\text{Gpc}^{-3}\text{yr}^{-1}$] as a function of redshift
\begin{equation}
    n^\text{F}(z) = n^\text{F}_0 \frac{(1+z)^{\alpha}}{1 + (\frac{1 + z}{C})^\beta},
\end{equation}
with $\alpha = 2.57$, $\beta = 5.83$ and $C = 3.36$; population-III binaries with
\begin{equation}
    n^\text{III}(z) = n^\text{III}_0 \frac{e^{a_\text{III}(z - z_\text{III})}}{b_\text{III} + a_\text{III}e^{(a_\text{III} + b_\text{III})(z - z_\text{III})}},
\end{equation}
with $a_\text{III} = 0.66$, $b_\text{III} = 0.30$ and $z_\text{III} = 11.6$; and primordial binaries with
\begin{equation}
    n^\text{P}(z) = n^\text{P}_0 \bigg(\frac{\tau(z)}{\tau_0}\bigg)^{-34/37},
\end{equation}
with $\tau_0$ being the current age of the Universe and $\tau(z)$ being the age of the Universe at redshift $z$.
Parameters are chosen as in~\cite{NgVitale2021}. Note, for the results in this section, we do not need to know the local merger rates $n^\text{F}_0$, $n^\text{III}_0$ and $n^\text{P}_0$ because we only require smaller example populations, but we require that they are drawn from reasonable redshift distributions. We draw the redshifts from the local merger rates $r(z)$ in units [$\text{yr}^{-1}$], which we obtain from the above equations using the relation:
\begin{equation}
    r(z) = n(z) \frac{4 \pi c D_\text{L}(z)^2}{H_0 (1+z)^3 E(z)},
\end{equation}
where $n(z)$ is the volumetric merger rate, $c$ is the speed of light, $D_\text{L}(z)$ is the luminosity distance, $H_0$ is the Hubble constant, $E(z)$ is the redshift scaling of the Hubble constant $H(z) = H_0 E(z)$.

The coherent SNR and null SNR for these populations are in Figure~\ref{fig:snrs}. We notice two important features. The first feature is that ET detects the entire sampled population of hundred-solar-mass field BBHs and population-III BBHs. Instead, since primordial black-hole mergers occur at redshifts $> 20$, detector-frame masses can be a few $1000~M_\odot$, and many of these signals either remain below the ET band or only show in band for a very short time with correspondingly low SNR.
%\hl{(what waveform model was used? did it include higher-order modes? this is to clarify since higher-order modes are for sure important for PBH observations.)}
The second feature is that a few high-SNR signals have null SNR exceeding the detection threshold. This is because, in reality, some residuals of a signal remain in the null stream due to finite time delays between the three components of ET. While these residuals can be significant, it will be possible to correct for these effects by considering time delays explicitly in the signal model when the signals surpass SNRs of a few thousands.

\begin{figure*}[!htb]
    \centering
    % \begin{subfigure}[b]{0.32\textwidth}
    %     \includegraphics[width=\textwidth]{plots_draft/snr_dist_null.pdf}
    %     %\caption{Field}
    %     \label{fig:j1017_1}
    % \end{subfigure}
    \begin{subfigure}[b]{0.49\textwidth}
        \includegraphics[width=\textwidth]{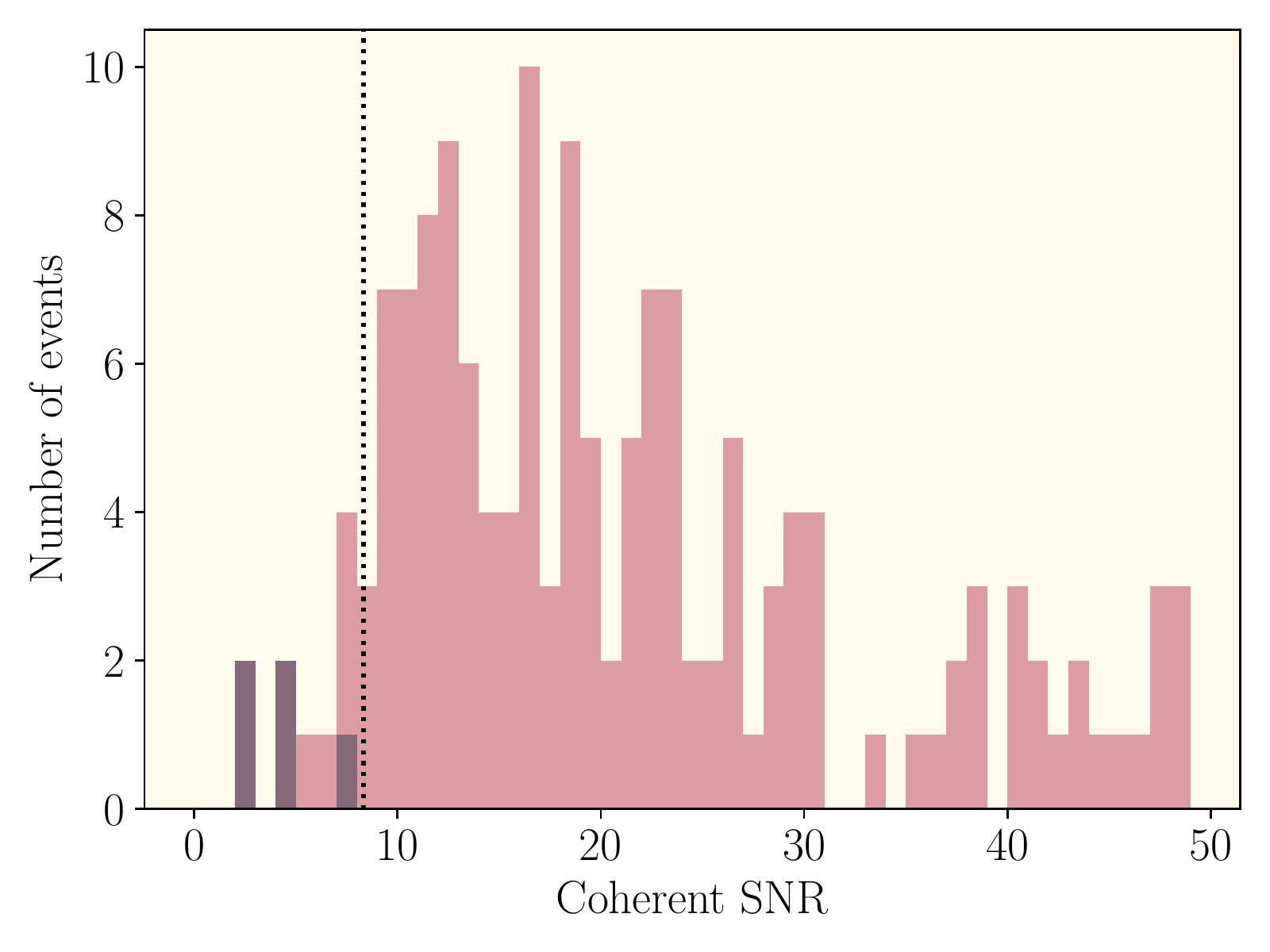}
        %\caption{Population-III}
        \label{fig:j1017_2}
    \end{subfigure}
    \begin{subfigure}[b]{0.49\textwidth}
        \includegraphics[width=\textwidth]{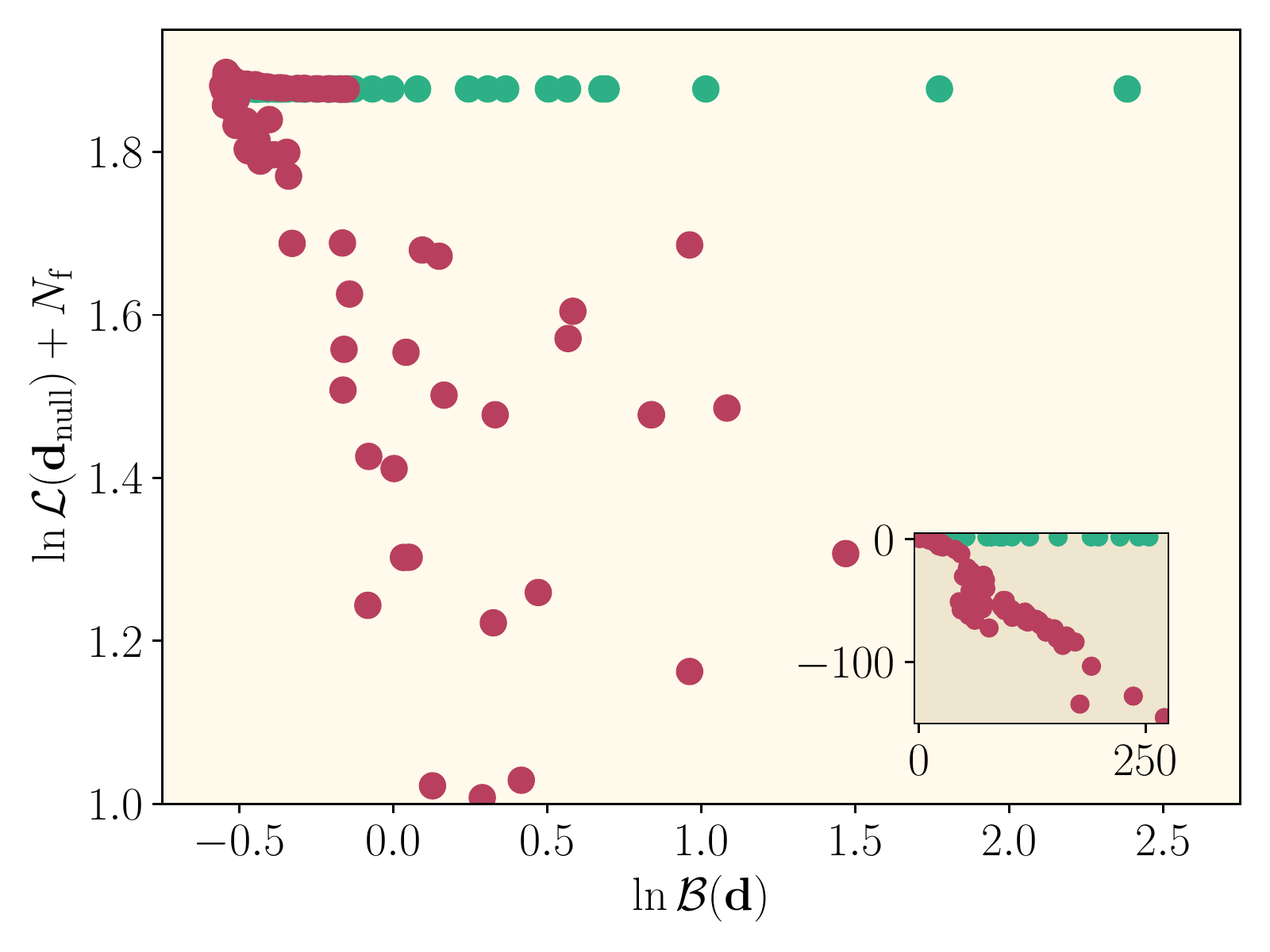}
        %\caption{Primordial}
        \label{fig:j1045_1}
    \end{subfigure}
    \vspace{-1.2\baselineskip} % to shorten space
    \caption{The effectiveness of glitch removal using the null stream. Left: the results of the null SNR based test described in Section~\ref{sec:glitches:nullsnr}. We show a distribution of coherent SNR for glitches, which we conservatively simulate as signals in only one of the three ET components. Red color represents all glitches and purple color represents glitches that survived the null stream veto.
    The veto removed all false positives down to the Gaussian noise levels.
    The dotted vertical line is the detection threshold which correspond to false alarm rate of $\text{yr}^{-1}$. 
    Right panel: the results of the null stream likelihood based test described in Section~\ref{sec:glitches:bayesian}. We plot the null stream likelihood $\ln \mathcal{L}(\bm{d}_\text{null})$ against the log Bayes factor $\ln \mathcal{B}$ for a signal hypothesis over the noise hypothesis in the interferometer data $\bm{d}$. Signals (green) and glitches (red) fall into two different areas, which allows to distinguish them starting from $\ln \mathcal{B}$ as low as $0$. The inset represents the results for an increased range of $\ln \mathcal{B}$.
    It is apparent that for Gaussian noise $\ln \mathcal{L}(\bm{d}_\text{null})$ are offset from $-N_\text{f}$ by $\approx 1.9$.
    This is because $N_\text{f}$ does not account for the loss of information during data-processing algorithms such as windowing.
    Given $N_\text{f}=1017$, the offset corresponds to $0.2 \%$ of data being lost.
    The distribution of $\ln \mathcal{L}(\bm{d}_\text{null})$ is shown in Figure~\ref{fig:glitchveto_lnl_dist}.
    }
    \label{fig:glitchveto}
\end{figure*}

Next, we demonstrate the performance of null SNR in mitigating glitches.
We simulate $200$ signals and $200$ glitches as $300~M_\odot$ detector-frame equal-mass black holes distributed uniformly between the luminosity distances of $10$ Mpc and $40$ Gpc, where glitches appear only in one of the three ET components.
This is the most conservative model for glitches in separate ET components which prohibits excising them based on morphology.
%\hl{(hmm. I would expect that a glitch with such characteristics should have an enormous SNR in the null stream... how is it that the null SNRs are so low for some of these glitches. I would expect that all glitch null SNRs are in the several thousands!)}
The mass and distance are chosen arbitrarily to roughly correspond to Figure~\ref{fig:glitchsignal}.
Next, we carry out a GW search and obtain coherent SNR and null SNR for the population, as we did for the astrophysical signals in the previous paragraph.
The distribution of coherent SNR we obtain for signals peaks at $\approx$ 100, and its minimum and maximum values are within an order of magnitude.
The distribution of null SNR for signals is centered at around $1.2$ with a maximum value of $3.0$.
For glitches, the distribution of coherent SNR is shown in red in the left panel of Figure~\ref{fig:glitchveto}.
The distribution of null SNR for glitches is centered around $50$ and reaches values as high as $1702$.
%The distribution of coherent SNR for glitches is in the right panel in Figure~\ref{fig:glitchveto}.
%In the left panel of Figure~\ref{fig:glitchveto}, we show the distribution of null SNR and coherent SNR for simulated astrophysical signals.
%In the left panel of Figure~\ref{fig:glitchveto}, we show coherent SNR for a population of glitches in blue.
%Even though the glitches are incoherent, the range of coherent SNR for our population reaches 50 due to excess power.
We veto glitches that have null SNR greater than $3.5$, suitable for a given number of transients.
The glitches that survived the veto are shown in purple in the left panel of Figure~\ref{fig:glitchveto}.
None of the survived glitches have coherent SNR exceeding the detection threshold of $8.3$, corresponding to the maximum value one can obtain with Gaussian noise in one year of observations.
We directly estimate this threshold using the assumption of a Gaussian noise background and number of trials given by typical BBH auto-correlation width $\mathcal{O}$(ms) and effective number of noise trials $\mathcal{O}$(1000).
Therefore, null stream has potential to excise all non-Gaussian outliers down to background expected from Gaussian noise alone.
The exception is the transient noise that couples simultaneously into each detector data stream with minimal time delay.

\subsection{\label{sec:glitches:bayesian} Likelihood-based approach}
%\hl{(the next two sections where in the previous section, but I think that the likelihood approach is the only one we want to show.)}
%We demonstrate the performance of null SNR in mitigating glitches. We simulate glitches to look exactly as $300~M_\odot$ detector-frame equal-mass black holes at $40$\,Mpc that appear only in one of the three ET detectors \hl{(hmm. I would expect that a glitch with such characteristics should have an enormous SNR in the null stream... how is it that the null SNRs are so low for some of these glitches. I would expect that all glitch null SNRs are in the several thousands!)}. Applying a simple null-SNR based glitch veto, the left panel in Figure \ref{fig:glitchveto} shows that five of the \hl{XXX} simulated glitches are not vetoed. 

An alternative implementation is to adopt the null likelihood described in Section~\ref{sec:likelihood}.
Unlike the null SNR which is based on fitting a waveform template to the null stream data, the likelihood-based approach does not depend on the signal model whatsoever.
%\hl{(this important point should be made clearer in this section. how does the null-SNR based veto depend on the signal model?)}
The null stream likelihood $\mathcal{L}(\bm{d}_\text{null})$ acts as Bayesian evidence for the noise-only hypothesis it represents. 

We simulate a population of signals and glitches, as in the previous section, based on the population of primordial black holes shown previously in Figure~\ref{fig:snrs}.
As shown in~\cite{NgChen2021}, high-order gravitational wave modes are important to identification and parameter estimation of such far away sources.
So, for this calculation, we use \texttt{IMRPhenomXPHM} waveforms~\cite{PrattenGarcia-Quiros2021}.
The results of the veto are shown in the right panel of Figure~\ref{fig:glitchveto}.
The horizontal axis represents signal loudness, $\ln \mathcal{B}(\bm{d})$, and the vertical axis represents deviations from Gaussian noise in the null stream, $\ln \mathcal{L}(\bm{d}_\text{null})$. With increasing loudness of both signals and glitches, null stream data for signals remains consistent with Gaussian noise, whereas louder glitches show stronger deviations from Gaussian noise in the null stream.
In our preliminary demonstration, the veto allows to distinguish signals from glitches  for both prominent transients with $\ln \mathcal{B} > 250$ and transients almost indistinguishable from the noise, with $\ln \mathcal{B} = 0$.
This way, null stream provides a powerful complementary opportunity to evaluate data quality.%\hl{(as I wrote on Slack, I think that the purpose of fig 4 should be to motivate a new SNR threshold. so, I would focus on low-SNR signals, like the PBH population. actually, why not taking exactly the same population already used in figure 3, c?)}

\section{\label{sec:psd} Estimating noise in the presence of overlapping signals}

The power spectral density (PSD) of detector noise forms an integral part of the detection and data-analysis pipelines of GW detectors, and details of how it is calculated are important to the results \cite{UsmanNitz2016,AdamsBuskulic2016,CannonCaudill2021}. In the 3G era, a large number of signals will be detected, and individual signals might be observed for an entire day or longer \cite{MaggioreVanDenBroeck2020,KalogeraSathyaprakash2021}. In this case, providing an estimate of the instrument-noise spectrum might be complicated due to GW signal contributions.
This could be a problem especially in the online analysis of GW data when waveforms are not accurately estimated yet to subtract the contribution of signals from the data.
Furthermore, stochastic GW searches can be improved if a precise noise PSD is provided \cite{ThraneRomano2013,RomanoCornish2017}.% as explained in greater detail in section \ref{sec:stochgw}.

To start with, we assess the problem of overlapping signals for the Einstein Telescope by evaluating strength, frequencies and time in the observing band of astrophysical signals at any given point in time.
We consider
%toy-model % JANPSD: we do not have specific numbers, so we say that we use a toy model
populations of merging BNSs, BBHs, and NS/BH binaries, based on predictions for rates of these events from population models and observations listed in GWTC-3~\cite{AbbottAbbott2019b,gwtc-3-pop}.
We do not discuss different sub-populations of these events like we did for BBHs in Section~\ref{sec:glitches}.
% We simulate a population of signals for four days and evaluate signal amplitudes and frequencies at the start of this period. We choose a period of four days to include contributions from all signals that might be at this point in time, give that some signals can last several days~\cite{}. % BORISPSD
%Average numbers of the simulated signals across multiple time segments are presented in Table~\ref{tab:pop}. % JANPSD
Next, we calculate $\tilde{h}_{+,\times}(f)$ [strain $\times$ s] using the frequency-domain
\texttt{IMRPhenomD} waveform~\cite{KhanHusa2016,HusaKhan2016}. % BORISPSD
%\texttt{TaylorF2} waveform~\cite{BuonannoIyer2009}.
We choose the frequency range for the waveform using the relation~\cite{AbbottAbbott2017c}
\begin{equation}
    f^{-8/3}(t) = \frac{(8 \pi)^{8/3}}{5} \bigg(\frac{G \mathcal{M}_\text{c}}{c}\bigg)^{5/3} (t - t_\text{c}),
\end{equation}
with the chirp mass $\mathcal{M}_\text{c}$ the chirp mass, the coalescence time $t_c$ coalescence time, and the physical constants $G$ and $c$.

\begin{figure*}
    \centering
    \begin{subfigure}[b]{0.45\textwidth}
        \includegraphics[width=\textwidth]{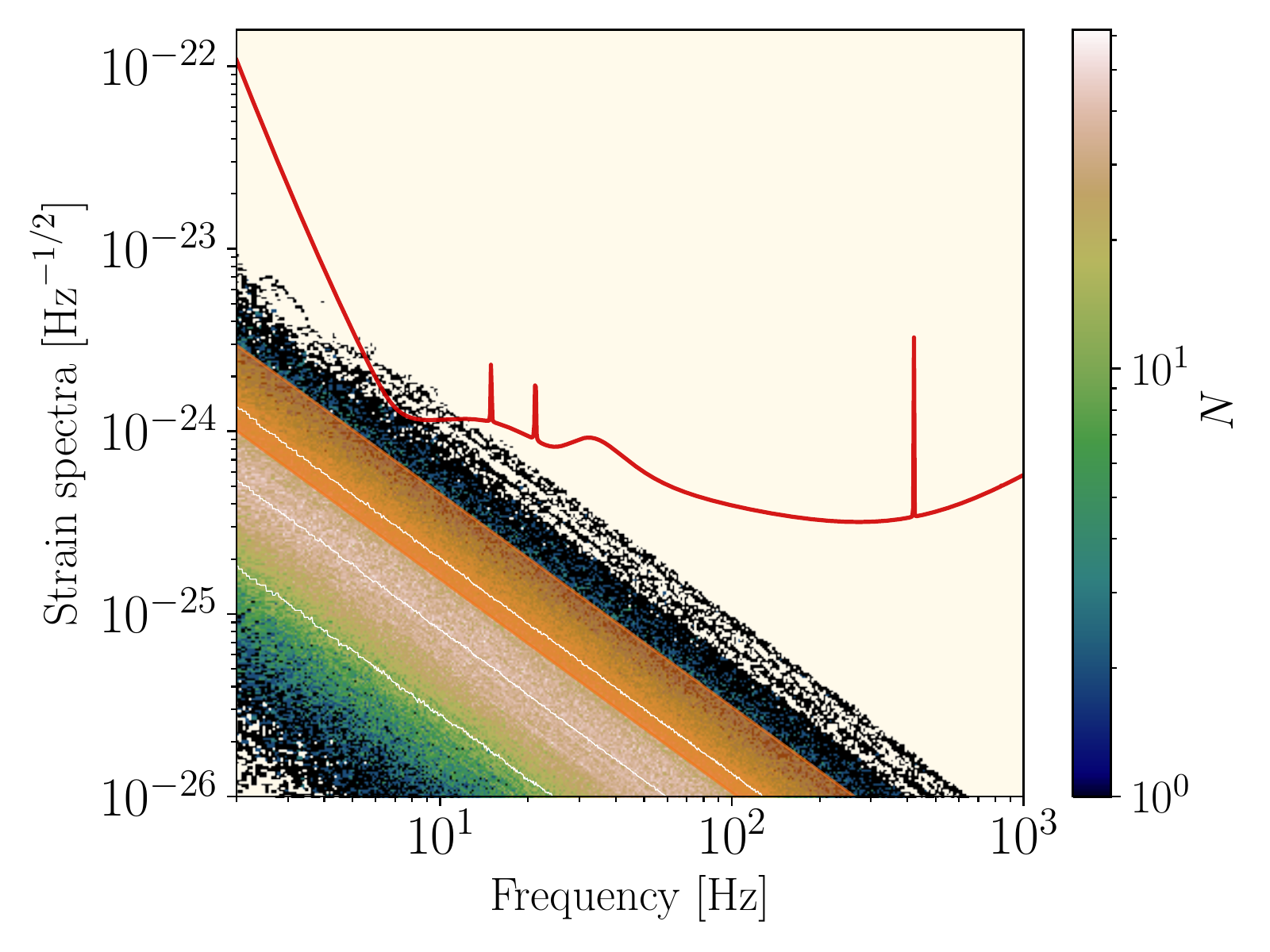}
      \caption{BNS, $T = 1$ minute}
        \label{fig:psd_cbc:3600_pmax}
    \end{subfigure}
    \begin{subfigure}[b]{0.45\textwidth}
        \includegraphics[width=\textwidth]{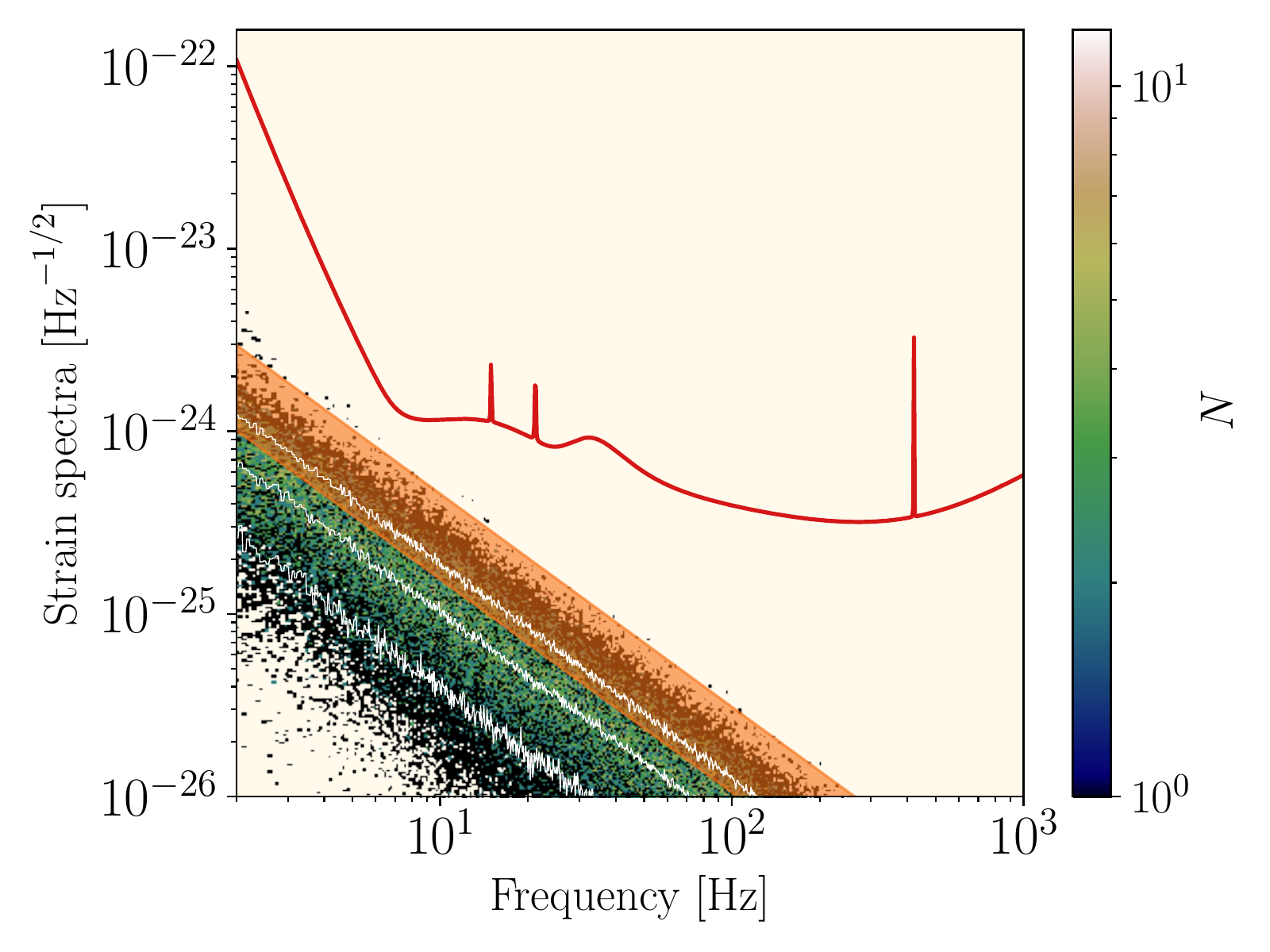}
      \caption{BNS, $T = 1$ day}
        \label{fig:psd_cbc:86400_pmax}
    \end{subfigure}
    \begin{subfigure}[b]{0.45\textwidth}
        \includegraphics[width=\textwidth]{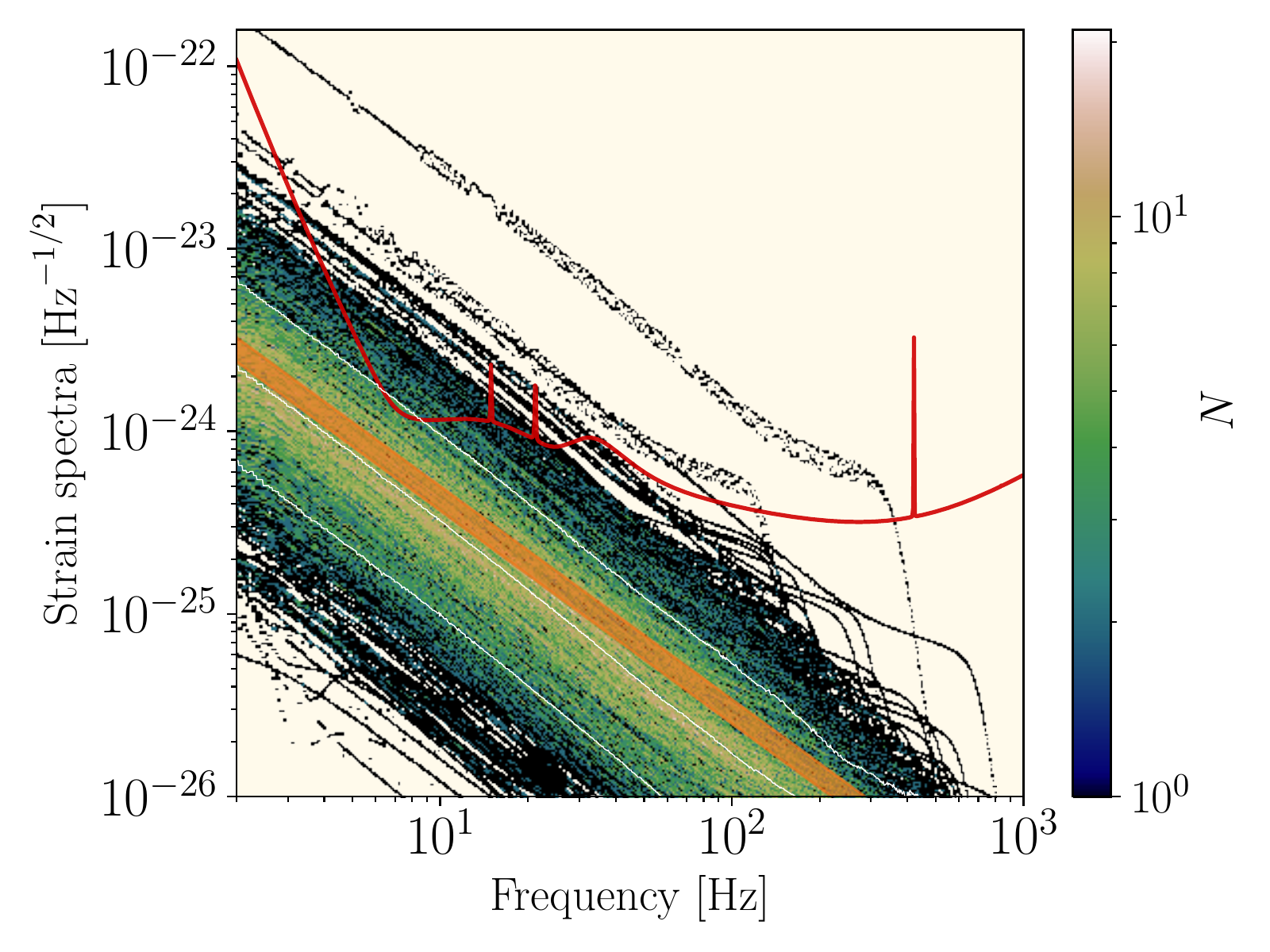}
      \caption{BBH, $T = 1$ minute}
        \label{fig:psd_cbc:3600_pmin}
    \end{subfigure}
    \begin{subfigure}[b]{0.45\textwidth}
        \includegraphics[width=\textwidth]{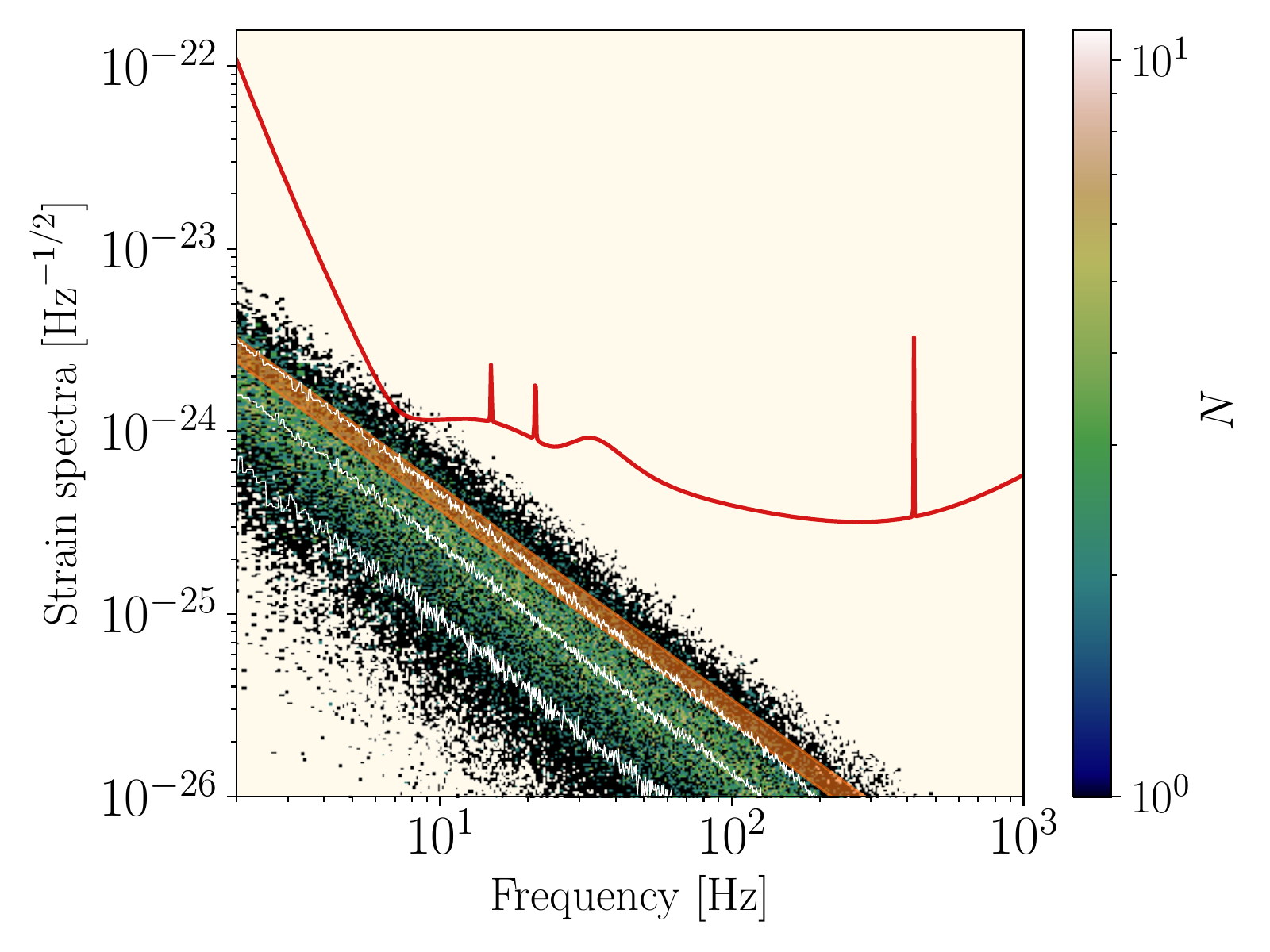}
      \caption{BBH, $T = 1$ day}
        \label{fig:psd_cbc:86400_pmin}
    \end{subfigure}
    \caption{Contributions of astrophysical signals to the measured power spectral density (PSD)\footnote{We present PSD in Figures~\ref{fig:psd_cbc} and~\ref{fig:nullpsd} as strain spectra, $\sqrt{P(f)}$, to follow conventions in the field.}. We plot histograms of single-segment PSD realizations for a BNS population (upper row) and a BBH population (lower row), where $N$ is the number of histogram counts at PSD bins for each frequency.
    The time segments used for the calculation of individual PSDs have a length of 1 minute with a total of 7200 realizations (left) and 24 hours with a total of 300 realizations (right).
    %The plots to the left show the 10th, 50th, and 90th percentiles of the PSD distributions as black-dashed curves.
    White lines show the 10th, 50th, and 90th percentiles of the PSD distributions.
    Around 2 BBH and 100 BNS contribute to 1-minute segments above 2 Hz, whereas 24-hour segments contain approximately 300 BBH and 2500 BNS.
    In orange bands, for comparison, we show ranges of the astrophysical background published in~\cite{AbbottAbbott2021b}. The red curve in all plots is the ET sensitivity.
    Occasionally, signal PSDs are larger than noise PSD, and thus PSD estimates of the data will contain significant contributions from compact binaries.
    Note, some of the reported uncertainties on merger rates $n_0$ from the LIGO-Virgo population study~\cite{gwtc-3-pop} cover a larger range of values and thus may yield even stronger contributions than presented in this figure.
    }
    \label{fig:psd_cbc}
\end{figure*}

Next, we evaluate the PSD of the signals with Equation~\ref{eq:psd_T} and determine the degree of their contribution to the noise PSD of the Einstein Telescope.
The results are provided in Figure~\ref{fig:psd_cbc}.
% The PSD can be defined as the limit~\cite{ThorneBlandford2017}
% \begin{equation}
% \label{eq:psd_T}
%     P(f) = \lim_{T \xrightarrow{} \infty} \frac{2 |\tilde{h}_T(f)|^2}{T},
% \end{equation}
% where $\tilde{h}_T(f)$ is the Fourier transform of a time series derived from $h(t)$ by setting its values 0 outside a time segment of length $T$.
For a comparison, time segments used to evaluate the noise PSD in current compact binary coalescence searches is $\sim 4~\text{s}$ for the low-latency \textsc{pycbc live} analysis~\cite{NitzDalCanton2018,DalCantonNitz2021} and typically $\sim8-16~\text{s}$ for archival \textsc{pycbc} analyses~\cite{UsmanNitz2016,DaviesDent2020}.
The final PSD estimate is obtained with based on a certain characteristic (e.g., mean or median) of the distribution of preliminary PSD estimates from several time segments segments within $\approx$ 1 minute~\cite{ChatziioannouHaster2019}.
%For future online searches in 3G era, $T$ can be \hl{...} ~\cite{}.
We show the PSD of the signal background observed for a minute and for a day in Figure~\ref{fig:psd_cbc}, for BBH and BNS signals.
As expected, the average PSD is the same for different $T$, which follows from Equation~\ref{eq:psd_T}.
This follows from ergodicity and stationarity of a stochastic process outlining the background.
We find that even though the signals are loud reaching high SNRs, signal contributions to the PSD are relatively low consistent with previous predictions \cite{RegimbauEvans2017,SachdevRegimbau2020}.
However, instrument noise PSD estimates can occasionally contain significant contributions from BBH signals and in rare cases also from BNS signals if short time segments are used to calculate the PSDs. If PSDs are calculated from 1-minute segments, less than 10\% of PSD estimates have a significant contribution from GW signals. The contribution of BNS and BBH signals to the PSD lie consistently below instrument noise for long PSD time segments of 24 hours duration. 

\begin{figure*}[!htb]
    \centering
    \begin{subfigure}[b]{0.49\textwidth}
        \includegraphics[width=\textwidth]{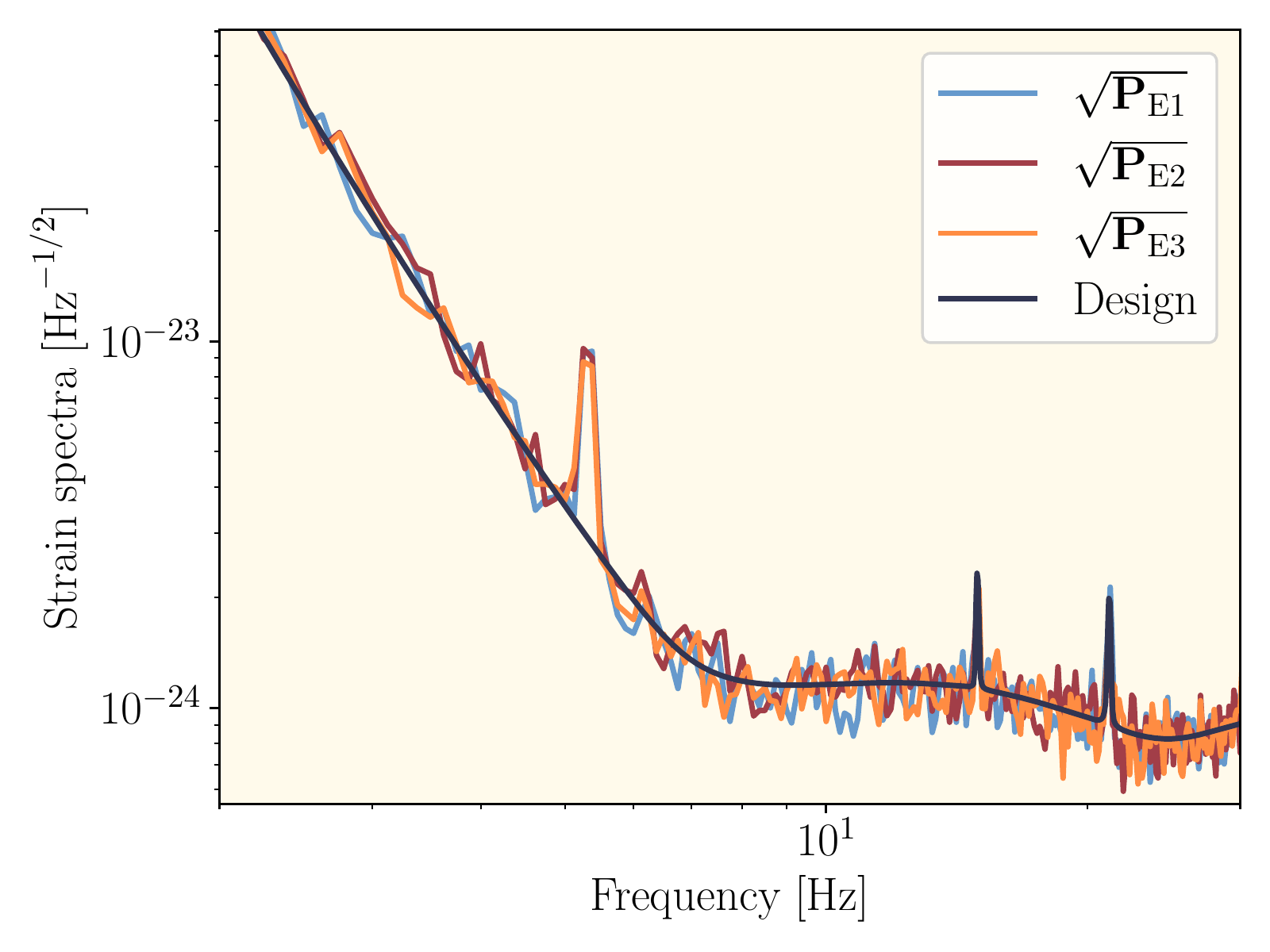}
        %\caption{Population-III}
        \label{fig:asd2-60}
    \end{subfigure}
    \begin{subfigure}[b]{0.49\textwidth}
        \includegraphics[width=\textwidth]{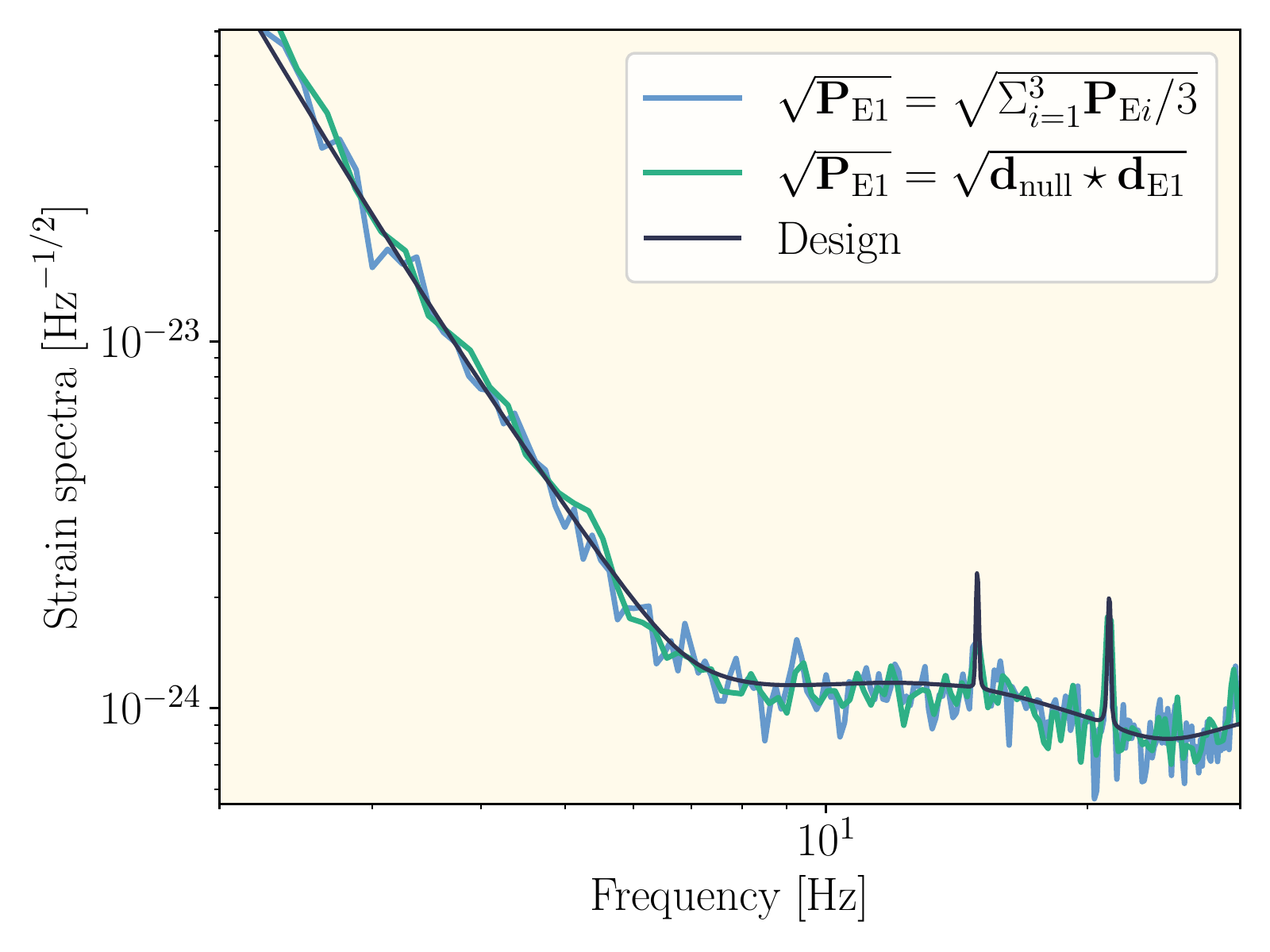}
        %\caption{Primordial}
        \label{fig:asd2-60-60}
    \end{subfigure}
    \vspace{-1.2\baselineskip} % to shorten space
    \caption{Evaluating noise power spectral density (PSD) in Einstein Telescope with the null stream. Left: total PSD in components of Einstein Telescope in presence of gravitational-wave signals from compact binaries.
    A BNS inspiral leaves a visible imprint on the inferred noise PSD at 5\,Hz where the measurement differs from the design sensitivity.
    PSD is estimated for a data spanning 128 s, which is why the BNS contribution is narrowband.
    Right: measurement of the PSD in one of the components based on the cross-power between the component and the null stream, as well as the linear relation between the null-stream PSD and the detector PSD.
    The contribution of the BNS signal is now removed from the PSD estimate.
    }
    \label{fig:nullpsd}
\end{figure*}

Next, we demonstrate estimation of noise spectra in the presence of a number of simultaneous signals in time series data. Normally, the noise PSD is calculated from time segments of data when GW signals are negligible, but it will be challenging to find these in ET, as we pointed out in the previous paragraph.
We simulate Gaussian noise in the three ET components based on the ET design sensitivity~\cite{ET2020} in 128 seconds of data.
%At this step, we show the measured PSD and the theoretical PSD in the left panel in Figure~\ref{fig:nullpsd}.
%With our population model, the time segment that we chose contains \hl{...} signals above 2 Hz, of which are \hl{...} BBH, \hl{...} BNS.
% and \hl{...} NSBH. % BORISPSD
Next, we simulate BBH and BNS in that data based on the population we used in this Section using \texttt{IMRPhenomPv2} waveforms~\cite{HannamSchmidt2014,KhanHusa2016,HusaKhan2016}.
Signals can add significant contributions to the detector PSD at low frequencies. An example is shown in the left panel in Figure~\ref{fig:nullpsd}.
%Thus, at ET design sensitivity it will be impossible to find a signal-free time segment between $\approx 2$ - $4$\hl{?} Hz to correctly evaluate the noise PSD directly from the detector strain output.
%Any of the simulated signals would remain undetected based on the incorrect overestimation of the noise.
Next, we show the recovery of the correct noise PSD in the first of the simulated ET components with the CSD-based method described by Equation~\ref{eq:noisepsd} in the right panel in Figure~\ref{fig:nullpsd}. We also demonstrate the relation between the PSD of the null stream and the PSD in the ET components, based on Equation~\ref{eq:psd}. As a result, we obtain the clean noise PSD for each of the ET components.

% The null stream provides an elegant method to provide unbiased estimates of instrument noise PSDs. Writing the data of the three components $k=1,2,3$ of the ET triangle as a sum over noise $n_k$ and signals $s_k$, $d_k=s_k+n_k$, and the null stream as the sum of the three components $\mathcal N=d_1+d_2+d_3=n_1+n_2+n_3$, and assuming that instrument noise in the three components is uncorrelated, then we obtain an estimate of the instrument noise PSD $S_k$ as a cross-power spectral density between the null stream and the data of each component,
% \begin{equation}
% S_k=\langle\mathcal N,d_k\rangle.
% \label{eq:noisepsd}
% \end{equation}
% This equation exploits the fact that signal content in the data of individual components of the ET detector do not find a correlated part in the null stream, however, the instrument noise of each component does. Therefore, clean instrument-noise PSD estimates could be given with low latency. 

% In the simplest case where sensitivities of detectors that constitute ET are the same, we solve Equation~\ref{eq:psd} for $P_{1,2,3}(f)$ and obtain the result in the right panel in Figure~\ref{fig:nullpsd}.
% The inferred detector noise PSD is consistent with the true noise PSD.
% \hl{Comment on the CSD method to infer noise PSD.}

\section{\label{sec:other} Other applications}

Systematic errors in detector calibration will leave residuals of GW signals in the null stream. Although one cannot directly obtain better absolute measurements of these errors with the ET null stream, one can still obtain better estimates on relative calibration errors between ET interferometers, and these can also help to understand systematic errors in the calibration procedure. The method proposed in~\cite{SchutzSathyaprakash2020} is to model residuals of GW signals in the null stream using a parameterization of calibration errors, which they call self-calibration. This method can be very effective if the number of detected signals is large; as expected in third-generation detectors like ET. They also pointed out that GW signals can in principle be used to transfer a precise calibration standard realized in only one ET interferometer at any frequency of the observation band to other frequencies and interferometers. It still leaves the challenge to realize a precise absolute calibration at least at one frequency with fractional calibration errors of about 0.1\% as needed for ET to fully exploit the information content in high-SNR signals. Challenges in the network null stream veto from calibration uncertainties are also discussed in~\cite{AjithHewitson2006}.

The model of the three-detector network null stream where signals cancel out is only applicable to GWs of General Relativity with two polarizations. Modifications and extensions of General Relativity include a possibility of additional, non-tensorial GW polarizations. For these models, GWs will leave residuals in the three-detector network null stream~\cite{WongPang2021}.
The situation is different for ET, where any field perturbations that produce the same length changes in two collocated arms of two different ET components are canceled in the null stream. E.g. $\Sigma_{i}(dx_i - dy_i)/L = 0$ because $dx_{1,2,3} = dy_{2,3,1}$ where $(dx_i,dy_i)$ are length perturbations in arms of a component $i$ and $L$ is the ET arm length.
Therefore, non-tensorial GW polarizations will also cancel in the ET null stream.

Furthermore, applications of the null stream outlined in Sections~\ref{sec:glitches} and~\ref{sec:psd} can be further generalized. The process of distinguishing glitches from compact binary signals can be extended to the case of continuous GWs and instrumental lines and long-duration noise transients, as well as to intermediate duration transients such that GW bursts and neutron star post-merger signals. This is due to the fact that null SNR in Section~\ref{sec:glitches:nullsnr} can include any signal model and the null stream likelihood term from Section~\ref{sec:glitches:bayesian} does not depend on a signal model and can be used as a data quality metric. 

Finally, the difference between the total PSD and the instrument-noise PSD represents the contribution of GW signals. It is demonstrated for compact binaries in~\cite{RegimbauDent2012} when assuming the same noise power spectra in ET components and employing Equation~\ref{eq:psd} for the noise PSD. 
The resulting average signal PSD had a characteristic spectral slope of the stochastic GW background, $P(f) \propto f^{-7/3}$~\cite{RenziniGoncharov2022}.
% JH: this would be strange since the characteristic slope of the astrophysical background is different, right?
Using Equation~\ref{eq:noisepsd} for the noise PSD allows us to extend the application to the case with different instrumental noise levels in constituent ET interferometers. Moreover, since the measurement of the astrophysical background and instrument-noise PSDs both inherit the same calibration errors when using Equation~\ref{eq:noisepsd}, the quality of the subtraction would be independent of calibration errors. Noise correlations between different ET interferometers can pose a limit here since the estimate of the instrument-noise PSD is biased by them.

\section{\label{sec:limitations} Limitations of null streams}

With all advantages of the null stream of the Einstein Telescope, there are a few caveats. It is possible that the interferometers that constitute ET will have significantly different sensitivities. Some null stream analyses such as glitch identification are limited by the interferometer with the lowest sensitivity. As shown in Figure~\ref{fig:snrs}, we also found that the ET null-stream analyses might potentially be limited for very loud signals, which can leave significant residuals in the null stream. The residuals are due to small time delays between ET components. Correct modeling of this effect requires abandoning the long-wavelength approximation. As a consequence, at this level of accuracy, the ET null stream becomes dependent on GW propagation direction, and this effect becomes stronger towards higher frequencies. Nevertheless, while this point deserves a more careful analysis, we expect that its impact on ET null-stream analyses is minor.

The null stream constructed from a network of observatories has additional limitations, but it can still be useful in data analysis. In the ET era, we expect a few hundred signals at any time in the observation band of ground-based GW detectors. After cancelling one signal in a network null stream, all of the remaining compact binary signals will still leave residuals there. The degree of contamination by compact binaries at low frequencies can be evaluated based on Figure~\ref{fig:psd_cbc}. Nevertheless, there is still room to apply network null streams to signals at high frequencies where temporal overlap is unlikely. Another limitation is that networks will have significantly different detector sensitivities by design and in practice.

\section{\label{sec:conclusion} Conclusion}

Thanks to its design, which enables a construction of the null stream, a synthetic data channel where all gravitational-wave (GW) signals cancel, the Einstein Telescope (ET) provides a solution to two important problems in GW analyses:
\begin{itemize}
    \item First, non-stationary incoherent noise artifacts, unlike GWs, immediately appear as outliers in the null stream, which allows to eliminate them without the need to perform glitch modeling or classification.
    We demonstrated how this can be achieved using the null-stream noise likelihood introduced in Section~\ref{sec:glitches:bayesian} and also using the null SNR, the coincident SNR in the null stream, first introduced in~\cite{HarryFairhurst2011}.
    The artifacts were eliminated down to the Gaussian noise background.
    %We find that the most distant BBHs that might originate from PBHs will not be confidently resolved.
    For primordial and the most distant Population-III BBH signals, which appear as weak, high-mass BBHs in the data, the benefit of the null stream will be particularly strong.
    %\RED{(isn't this simply because of sub-threshold SNR? I mean, figure 3,c is simply a statement of SNR and null SNR, and in this plot, there is nothing about glitches, right? putting this together with figure 4, right, I would say that also for PBH binaries, we can remove all influence from glitches down to Gaussian noise. The fact that the most distant PBHs cannot be detected is not relevant for this paper unless connected to the glitches.)}
    \item Second, ET will be able to provide unbiased estimates of noise-power spectra even for segments of data where multiple loud signals are present. We show this will often be the case with the ET design sensitivity at low frequencies, and recent predictions for compact-binary merger rates~\cite{gwtc-3-pop}. 
\end{itemize}

Our analysis only considered so-called blip glitches, which we modeled as short-duration BBHs.
Discrimination between signals and glitches as outlined in this article can be extended to the case of other GW signals, including bursts from cosmic strings or supernovae, neutron star post-merger signals, continuous waves, and other new signals that 3G detectors might confidently detect for the first time. In particular, the discrimination will be relevant for signals described only by phenomenological models or for signals without a clear model. 

Estimation of noise spectra using the ET null stream provides advantages compared to both the off-source PSD estimation using time segments adjacent to signals and on-source methods that evaluate PSD simultaneously with signals, e.g. \textsc{bayesline}~\cite{LittenbergCornish2015} now integrated in \textsc{bayeswave}~\cite{CornishLittenberg2015,CornishLittenberg2021}. For ET, finding signal-free data segments needed for the off-source methods might be challenging, especially for online analyses, due to the abundance of signals in band at any given time.
Whereas on-source methods perceive astrophysical contamination as yet another component of the noise, so they would still overestimate the instrumental noise PSD. Moreover, on-source methods still have more dependence on the PSD model compared to the ET null stream, and too-flexible models increase the measurement uncertainty of GW parameters.
PSD estimation with ET null stream in the presence of glitches has not been explicitly demonstrated in this work, but we foresee it will be possible as well. Time segments of data where glitches were identified can be either removed from the analysis or left with glitch being modeled or subtracted.

It is worth keeping in mind that we provided the results under certain assumptions we outlined in Section~\ref{sec:method}. For example, we do not quantify the effects of noise correlations between ET interferometers on the null stream, which is an important aspect that needs to be investigated. We close our remarks by pointing out that the generic network null streams have significant limitations compared to the ET null stream. The main advantage of the ET null stream is that it does not depend on propagation directions of GWs, and it is therefore also valid in the presence of multiple GW signals.

\section*{Acknowledgments}

%The code to reproduce our calculations is available at~\hyperlink{https://github.com/bvgoncharov/null_stream_2021/}{github.com/bvgoncharov/null\_stream\_2021/}.
We thank Jacopo Tissino for helpful comments.
We make use of \textsc{pycbc}~\cite{NitzHarry2022}, \textsc{bilby}~\cite{AshtonHubner2019}, and \textsc{gwfish}~\cite{HarmsDupletsa2022}. BG is supported by the Italian Ministry of Education, University and Research within the PRIN 2017 Research Program Framework, n. 2017SYRTCN.

% Can add in further derived cases that are the result of specific technical 
% capabilities
%   High redshift (and massive source) detection
%           impact for PBHs, pop III, 
%           easier to get background down to Gaussian wall
%
%   more easily enables pre-merger alerts / rapid accurate sky localizations
%           [result of confusion noise suppression -> easier to estimate PSD]
%           larger impact due to confusion noise at low frequencies

\bibliographystyle{apsrev}
\bibliography{mybib,collabdocs}{}

\appendix*

\begin{figure*}[!htb]
    \centering
    \includegraphics[width=\textwidth]{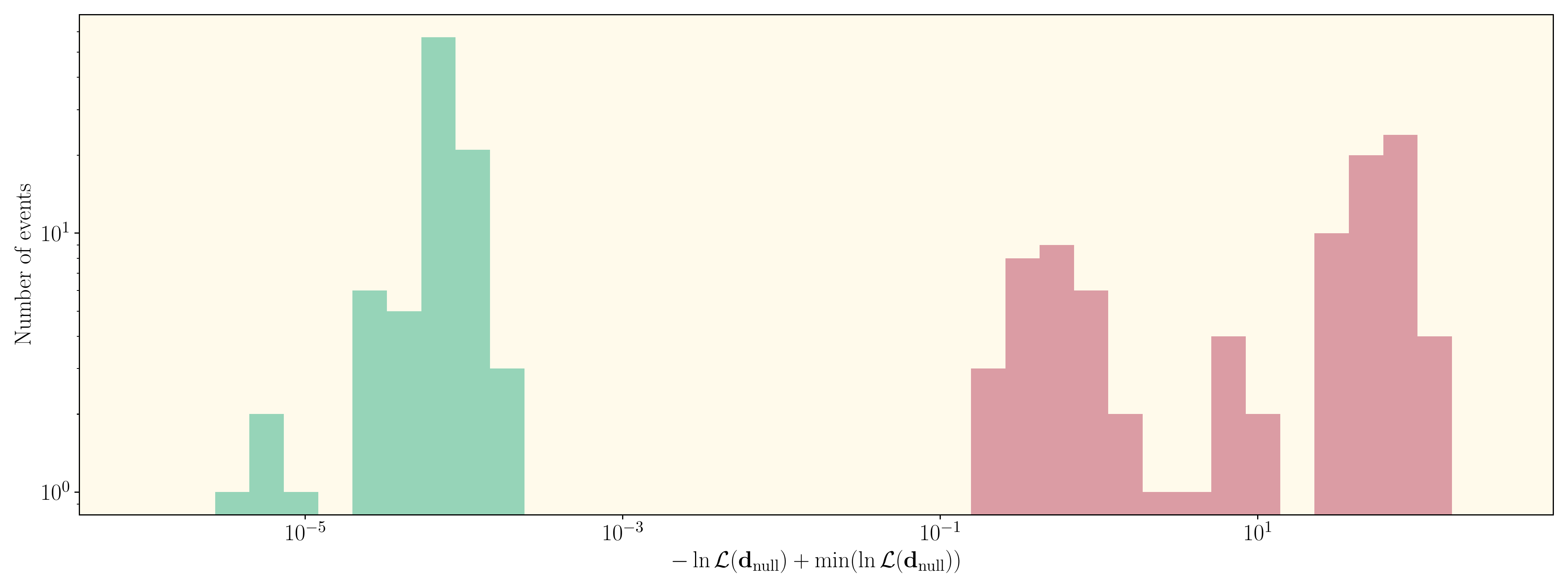}
    \caption{The distribution of the null stream likelihood, $\ln \mathcal{L}(\bm{d}_\text{null})$, referenced to the minimum value, for the BBH signals (green) and glitches that we conservatively model as incoherent BBH signals (red). The data corresponds to the vertical axis in Figure~\ref{fig:glitchveto}, excluding the values consistent with pure Gaussian noise, such that $\ln \mathcal{B}(\bm{d})<0$.
    The distribution of glitches is separated from that of signals by several orders of magnitude of the width of the signal distribution, which proves the null stream likelihood to be an effective glitch veto.
    }
    \label{fig:glitchveto_lnl_dist}
\end{figure*}

\end{document}